\documentclass[structabstract]{aa}
\pdfoutput=1
\usepackage{txfonts}
\usepackage{graphicx}
\usepackage{natbib,twoopt}
 \bibpunct{(}{)}{;}{a}{}{,}    

\def\textsf{$\textsf M$}

\def\textsf{$\textsf M$}

\def\o5{[\ion{O}{iii}]$\lambda$5007}
\def\03{[\ion{O}{ii}]$\lambda$3727}
\def\xo5{[\ion{O}{iii}]$\lambda$5007~}
\def\xo3{[\ion{O}{ii}]$\lambda$3727~}

\def\sma{$\textsf M_\odot$}

\def\sma{$\cal M_\odot$}
\def\mst{$\cal M_\star$}
\def\mass{$\cal M$}

\begin{document}

\title{Escape of Lyman continuum radiation from local galaxies}
\subtitle{Detection of leakage from the young starburst Tol 1247-232}


\author{Elisabet Leitet\inst{1}
\and Nils Bergvall\inst{1}
\and Matthew Hayes\inst{2,3}
\and Staffan Linn{\'e}\inst{1}
\and Erik Zackrisson\inst{4}}

\institute{{Department of Physics and Astronomy, Uppsala University, Box 515, SE-751 20 Uppsala,
Sweden}\\
\email{elisabet.leitet@physics.uu.se, nils.bergvall@physics.uu.se, staffan.linne@gmail.com}
\and {Universit\'e de Toulouse, UPS-OMP, IRAP, Toulouse, France}
\and {CNRS, IRAP, 14 avenue Edouard Belin, F-31400 Toulouse, France}\\
\email{matthew.hayes@irap.omp.eu}
\and{Department of Astronomy, Stockholm University, Oscar Klein Center, AlbaNova, Stockholm SE-106 91, Sweden}
\email{ez@astro.su.se}}

\date{Accepted}

\abstract {The escape fraction of hydrogen ionizing photons ($f_{esc}$) from galaxies has been suggested to be evolving with time, but the picture is far from clear. While evidence for significant escape fractions has been found at high redshifts in several studies, the picture looks different in the more nearby universe. }
{Here, we apply a new background subtraction routine on archival data from the Far Ultraviolet Spectroscopic Explorer (FUSE), in order to study local galaxies in search for possible Lyman Continuum (LyC) leakage. In the process, for the first time a stacked spectrum in the LyC is produced  for local galaxies. With this small sample, we also make a more tentative approach to look for possible correlations between $f_{esc}$ and physical parameters such  as internal absorption E(B-V)$_i$, mass, \ion{H}{I} mass, specific star formation rate (SSFR), metallicity, and Ly$\alpha$ emission.}
{Eight star forming galaxies with redshifts z $>$ 0.015 from the FUSE archive were re-examined. Also, a sub-sample of an additional four galaxies  with lower redshifts were included, for which the escape fraction was estimated from residual flux in the low ionization interstellar \ion{C}{ii} $\lambda$1036 {\AA} line.}
{Out of the eight galaxies, only one was found to have significant LyC leakage, Tol 1247-232 (S/N=5.2). This is the second detection of a leaking galaxy in the local universe. Using the first case, Haro 11, as a calibrator, we find an intrinsic Lyman break amplitude for starbursts at this young age  of ($f_{1500\AA}/f_{900\AA})_\mathrm{int}$=1.5$^{+0.6}_{-0.5}$, which gives an absolute escape fraction for Tol 1247-232 of $f_{esc}$ = 2.4$^{+0.9}_{-0.8}$ $\%$. The stacked sample show an excess in the LyC with $f_{esc}$ = 1.4$^{+0.6}_{-0.5}$ $\%$, but we note that there might be important selection biases involved as the galaxies originally were handpicked for their star forming qualities. With the small sample, we suggest a possible trend for higher $f_{esc}$ with lower mass and with enhanced SSFR.  None of the galaxies with high values of  E(B-V)$_i$ were found to show any sign of leakage.}
{}




\maketitle

\section{Introduction}
Recently, there have been several detections at high redshifts of galaxies leaking hydrogen ionizing photons. Even though it seems likely that galaxies played an important role in the reionization of the universe (e.g. \citealt{2010ApJ...710.1239R,2011arXiv1105.2038B}), we are not yet detecting the galaxies responsible since these would be well below current detection thresholds \citep{2010MNRAS.409..855B,2010MNRAS.403..960M}. While we cannot directly measure  the amount of  Lyman Continuum (LyC) photons that escape from the reionization-epoch galaxies, observations of low- to intermediate-redshift objects have given tentative evidence for an evolving escape fraction (e.g. \citealt{2006MNRAS.371L...1I,2010ApJ...723..241S}). There have been several detections of individual LyC leaking galaxies at z $\sim$ 3  (\citealt{2012arXiv1210.2393N,2011ApJ...736...18N,2010ApJ...725.1011V,2009ApJ...692.1287I}), although foreground contaminants are expected to account for some of those. Also, a few stacked samples at the same redshift show LyC leakage (e.g.  \citealt{2001ApJ...546..665S,2011ApJ...736...18N}), while yet others  seems to contradict an evolving escape fraction scenario by only yielding upper limits of a few percent at z $\sim$ 3--4 \citep{2010ApJ...725.1011V,2011ApJ...736...41B}. The different methods applied can however have an impact on these results, something which is further discussed below.

At lower redshifts, the picture looks different. The investigations targeting z $<$ 3 galaxies have all, except one, resulted in non-detections and upper limits on the escape fraction.  For galaxies at  intermediate redshifts, z $\sim$ 0.7-1.4 \citep{2010ApJ...720..465B,2010ApJ...723..241S,2007ApJ...668...62S,2009ApJ...692.1476C},  upper limits of only 1-2 $\%$ have been derived for stacked samples, and there are no individual detections. In the local universe, several attempts have been made to measure escaping ionizing photons from galaxies using space based telescopes -- first with The Hopkins Ultraviolet Telescope \citep{1995ApJ...454L..19L,1997ApJ...481L..31H}, and later with FUSE \citep{2001A&A...375..805D,2006A&A...448..513B,2009ApJS..181..272G}. The difficulty in handling the background and other noise sources such as airglow and scattered light has however made these data hard to interpret. \citet{2001A&A...375..805D} derived an upper limit of $f_{esc}$ $<$ 6.2 $\%$ for Mrk 54, and a few years later  \citet{2006A&A...448..513B} made the first detection for the blue compact galaxy Haro 11. This result could however not be confirmed by \citet{2007ApJ...668..891G}, but by developing a new method to approach the sensitive FUSE background subtraction, the leakage was confirmed by us with $f_{esc}$=3.3$\pm$0.7$\%$ (\citealt{2011A&A...532A.107L}, hereafter L11). Haro 11 is thus, up till now, the only known LyC leaking galaxy below z $<$ 3.

The lack of LyC leaking galaxies at lower redshifts is perhaps not so unexpected. Most models indicate that the low mass galaxies abundant in the early universe, are more likely to have higher escape fractions than the larger systems that eventually emerged via  hierarchical structure formation (e.g. \citealt{2010ApJ...710.1239R,2011MNRAS.412..411Y}).  Also, in the early universe, the existence of  metal-free stars (so-called population III stars; hereafter pop III) (e.g. \citealt{2003A&A...397..527S,2009MNRAS.399...37J,2011ApJ...740...13Z}), a lower dust content, a more top-heavy initial mass function (IMF), smaller galaxy sizes and runaway stars \citep{2012ApJ...755..123C} are factors that could affect the overall escape fraction from galaxies. However, there are several important differences with the methods used to find leaking galaxies at different redshifts. At high redshifts, z $\sim$ 3, the detections have mainly been made with field observations using narrowband imaging or spectroscopy just below the Lyman limit. For several of these detections the $f_{1500}$/$f_{LyC}$ ratio has implied escape fractions above unity for any SED model. The cause of this is not yet understood, but can possibly be explained by bound-free emission from the nebular gas heated by pop III stars causing a "Lyman bump" just below the Lyman limit as discussed in \citet{2010MNRAS.401.1325I} and \citet{2011MNRAS.411.2336I}. No detections have been made using  imaging at intermediate redshifts, but the investigations at z $\sim$ 1 utilize broader filters well below the Lyman limit. Also, the z $\sim$ 3 non-detections by  \citet{2010ApJ...725.1011V} and \citet{2011ApJ...736...41B} are broadband observations probing somewhat bluer wavelengths, and might thus perhaps not be comparable to the measurements just below the Lyman limit for the z $\sim$ 3 detections. In the local universe, fewer targets have been observed, and it is not possible to observe large multi-object fields at these short wavelengths. The targets have instead traditionally been singled out for their star forming properties  such as a strong H$\alpha$ emission, a selection criterion that might be counterproductive as it would give preference to ionization bounded H II regions with low escape fractions.


We would perhaps not predict high escape fractions from local galaxies, and in many regards they cannot be considered as good templates for the reionization epoch population. In the case of Haro 11, the 3 $\%$ leakage is indeed significantly lower than what has been derived for z $\sim$3 leakers. Still, local galaxies can give us an unique insight about the physics that regulates the leakage of ionizing radiation from galaxies, and give valuable information that can only be obtained from detailed studies of such nearby objects. Therefore, the FUSE archive still presents a unique legacy of local galaxies observed below the Lyman limit, even though the reduction pipeline, CalFUSE, has had some issues associated with the background subtraction for faint targets. Here, using the method presented in L11, we have taken a closer look at the  remaining FUSE targets at sufficient redshifts, such that the LyC can be measured above the Milky Way Lyman limit (Sections ~\ref{LyC} and ~\ref{fesc}). For the first time, a stacked LyC spectrum is produced for local-universe galaxies  (Section ~\ref{stack}). In Section ~\ref{Cii} we estimate the escape fraction from residual flux in the low ionization interstellar line  \ion{C}{ii} $\lambda$1036 {\AA}, following the work by \citet{2001ApJ...558...56H}.

In Section ~\ref{corr}, we  take on a more  tentative approach for this small sample of galaxies, in search for correlations between the escape fraction of ionizing photons to other galaxy parameters, such as intrinsic absorption E(B-V)$_i$, mass, \ion{H}{I} mass, specific star formation rate (SSFR),  dust and metallicity. We also want to investigate if we can find any relation between Ly$\alpha$ emission and LyC leakage as has been found at higher redshifts (e.g. \citealt{2011ApJ...736...18N}). 

A  flat $\Lambda$CDM cosmology with H$_0$=70 km s$^{-1}$ Mpc$^{-1}$, $\Omega$$_m$=0.3, and $\Omega$$_0$=0.7 was adopted throughout the article. All magnitudes are given in the AB system, and flux densities in f$_{\lambda}$.

\begin{table*}[htb!]
	\centering
	\caption{Galaxy properties.  (1) Name of the target galaxy, (2) Other name, (3) The classification, with documented Wolf-Rayet property in parenthesis,  
	(4) The absolute B-magnitude, 
	(5) The redshift,  (6) The redshifted Lyman limit [\AA], (7) The Galactic extinction, (8) The gas phase oxygen abundance as 12+log(O/H), 
	(9) The H$\alpha$ equivalent width [\AA], (10) The IUE UV continuum slope after correction for Galactic extinction, (11) The intrinsic absorption.}
	\begin{tabular}{l l l l l l l l l l l}
	\multicolumn{11}{c}{}\\
	\hline \hline
\scriptsize{Name} & \scriptsize{Other name} & \scriptsize{Classification}   &  \scriptsize{M$_B$} & \scriptsize{Redshift}& \scriptsize{Ly-limit} & \scriptsize{E(B-V)$_G$} & \scriptsize{12+log(O/H)} & \scriptsize{EW(H$\alpha$)} & \scriptsize{IUE $\beta$} & \scriptsize{E(B-V)$_i$} \\ 
	\scriptsize{(1)} & \scriptsize{(2)} & \scriptsize{(3)} & \scriptsize{(4)} & \scriptsize{(5)$^{b}$} & \scriptsize{(6) [\AA]} & \scriptsize{(7)$^{x}$} &  \scriptsize{(8)} & \scriptsize{(9) [\AA]} & \scriptsize{(10)$^{y}$} & \scriptsize{(11)$^{y}$} \\
	\hline 
	Haro 11 & ESO 350-IG 038 & BCG$^{v}$(WR$^{v}$) & -20.3$^{v}$ & 0.0206 & 930.3 & 0.012 & 7.9$^{i}$ & 704$^{i}$ & -1.38 & 0.17 \\ 
	VV 114 & ESO 541-IG 023 & LIRG$^{b}$  & -20.0$^{b}$ & 0.0201 & 929.8 & 0.016 &8.6$^{g}$ & 147.2$^{ac}$ & -- & 0.34$^{z}$  \\ 
	MRK 0357 & PGC 005010 & SB$^{t}$  & -21.4$^{h}$ & 0.0528 & 959.6 & 0.066 & 8.5$^{f}$ & 242$^{h}$  &  -1.19 & 0.21 \\
	$^{\dagger}$SBS 0335-052 & SBS 0335-052E & BCG$^{w}$(WR$^{a}$) & -17.0$^{i}$ & 0.0135 & 923.8 & 0.047 & 7.3$^{aa}$ & 1434$^{i}$ & -2.09 & 0.00 \\
	Tol 0440-381 & SCHG 0440-381 & H II$^{b}$(WR?$^{a}$)  & -20.2$^{b}$ & 0.0409 & 948.8 & 0.016 & 8.2$^{e}$ &  208$^c$ & -0.95 & 0.27 \\ 
	IRAS 08339+6517 & PGC 024283  & H II$^{w}$(WR$^{j}$) &  -21.0$^{i}$ &  0.0191 & 928.9 & 0.092 & 8.7$^{ab}$ & 199$^{i}$ & -1.74 & 0.08 \\
	$^{\dagger}$NGC 3991 & Haro 5 & BCG$^{k}$ & -20.7$^{k}$ & 0.0106 & 921.2 & 0.022 & 8.6$^{n}$ & 155$^{o}$ &  -1.74 & 0.08 \\
	$^{\dagger}$NGC 4194 &  UGC 07241  & BCG$^{t}$ &  -20.7$^{k}$ & 0.0083 & 919.1 & 0.016 & 8.7$^{l}$ & 109$^{p}$ &  -0.08 & 0.47 \\
	Tol 1247-232 & EC 12476-2317  & H II$^{b}$(WR$^{a}$) &   -21.0$^{b}$ & 0.0480 & 955.3 & 0.089 & 8.1$^d$ &  530$^c$ & -1.22 & 0.21 \\
	MRK 54  & MCG +06-28-044 & BCG$^{k}$ &  -22.2$^{k}$ & 0.0449 &  952.4 & 0.015 & 8.6$^{k}$  & -- & -1.58 & 0.12 \\
	MRK 499 & IZw 166  & BCG$^{k}$ & -21.1$^{k}$ & 0.0260 & 935.2 & 0.016 & 8.5$^{f}$ & 34.8$^{ac}$ &  -0.55 & 0.36 \\
	$^{\dagger}$ESO 338-IG04 & Tol 1924-416  & BCG$^{s}$ (WR$^{a}$) &  -18.9$^{s}$ & 0.0095 & 920.2 & 0.087 & 7.9$^{i}$ &  570$^{i}$ & -1.95 & 0.06 \\
	ESO 185-IG013 & PGC 063618 & BCG$^{k}$  & -20.4$^{k}$ & 0.0186 & 928.5 & 0.054 & 8.5$^{t}$ & -- &  -1.58 & 0.12 \\
	\hline
	\end{tabular}
	\\
	\scriptsize{$^{\dagger}$ For this galaxy the LyC leakage is estimated from residual flux in the \ion{C}{II} $\lambda$1036 {\AA} line only. \\
	$^{a}$ \citet{1999A&AS..136...35S}, $^{b}$ The NASA/IPAC Extragalactic Database (NED), $^{c}$ \citet{1991A&AS...91..285T}, $^d$ \citet{1993MNRAS.260....3T}, $^{e}$ \citet{1986MNRAS.223..811C}, $^{f}$ \citet{1996ApJ...466..831G}, $^{g}$ \citet{2009ApJS..181..272G} , $^{h}$ \citet{2005AJ....129.1863K}, $^{i}$ \citet{2009AJ....138..923O}, $^{j}$ \citet{1991ApJ...377..115C}, $^{k}$ \citet{1993ApJS...86....5K} , $^{l}$ \citet{2008ApJ...678..804E}, $^{m}$ \citet{1998ApJ...503..646H}, $^{n}$ \citet{1988A&A...205...41A}, $^{o}$ \citet{1992ApJ...388..310K}, $^{p}$ SDSS, $^{q}$ \citet{2004AJ....127.1360W},  $^{r}$ \citet{1996ApJS..107..521V}, $^{s}$ \citet{2002A&A...390..891B},  $^{t}$ \citet{1994ApJ...429..582C}, $^{v}$ \citet{2006A&A...448..513B}, 
	$^{w}$ \citet{2011AJ....141...37L}, $^{x}$ \citet{1998ApJ...500..525S}, $^{y}$ this work, $^{z}$  L(UV)/(L(UV)+L(IR)) from \citet{2009ApJS..181..272G} converted to E(B-V)$_i$, $^{aa}$ \citet{1999ApJ...511..639I},  $^{ab}$ \citet{1998ApJ...495..698G}, $^{ac}$ \citet{2006ApJS..164...81M}.}
	\label{table:tfour}
\end{table*}	

\section{The targets}
\subsection{Sample selection}
The FUSE database at the Multimission Archive at STScI (MAST) was searched for star forming galaxies (no active galactic nuclei, AGN).  The  selection criterion was basically the redshift, such that the signal in both detector segments covering the LyC wavelengths (SiC 1B and SiC 2A) could be compared over a  minimum of 12 pixels. To gain the signal-to-noise ratio (S/N) needed for the background model,  our data does not make use of the full FUSE resolution (Section~\ref{sec:model}). Therefore, since the Milky Way (MW) Lyman lines are densely distributed below 920 {\AA}, our sample for direct LyC measurements require redshifts of z $>$ 0.015. We also required a S/N $>$ 4 in the stellar continuum at rest wavelength $\sim$945 {\AA} measured on orbital night data exclusively. It has been shown in several previous articles that using night data only is a requirement to constrain the background of scattered light when working with low S/N FUSE data (e.g. \citealt{2004ApJ...605..631Z,2006ApJ...647..328D,2006A&A...455...91F}). The sample thus consists of 8 galaxies with redshifts 0.017 $<$ z $<$ 0.053. Further, we have also included Haro 11 in the sample, although the LyC leakage was analyzed already in L11. All but one galaxy, VV 114, have also been observed in Ly$\alpha$ by the  International Ultraviolet Explorer (IUE).

Since one of the aims in this work was to look for possible correlations between LyC and Ly$\alpha$ emission, a sub-sample was included to improve the statistics. These galaxies were selected for being star forming (no AGNs), and for being observed with both FUSE and IUE. Also, the redshift had to be high enough such that the intrinsic Ly$\alpha$ line could be de-blended from the strong geocoronal Ly$\alpha$ emission line in the IUE data. The LyC escape fraction was  for this sample estimated by measuring the residual flux in the \ion{C}{ii} $\lambda$1036 {\AA} line. This sub-sample consists of 4 galaxies with redshifts 0.008$<$z$<$0.013, and the final selection hence consists of 13 galaxies. 

The  galaxies are presented  in Table~\ref{table:tfour}.

\subsection{The properties of the galaxies} \label{iue}
More than half of the sample was originally observed by FUSE with the primary science aim to look for LyC leakage. However, few results have been published, and except for the previously mentioned Haro 11 only upper limits on $f_{esc}$ have been derived \citep{2001A&A...375..805D,2009ApJS..181..272G}. 

The sample is a mixture of galaxies classified as starbursts (SB), blue compact galaxies (BCG), H II galaxies (H II), luminous infrared galaxies (LIRG), and many of them are interacting systems.  Wolf Rayet (WR) features are documented  for half of the sample, which implies the presence of stars with ages  $\leq$10 Myr, so these galaxies must presently be undergoing a starburst phase or one was ended only very recently. Several of the targets display high H$\alpha$ equivalent widths, although the numbers presented in Table~\ref{table:tfour} in many cases were derived using a small aperture on the central region, and cannot directly be compared with other parameters measured from  the larger UV apertures. The metallicity of the galaxies range from very low to solar, with oxygen abundances in the range 12+log(O/H)=7.3 -- 8.7, but these values are also typically derived for the central regions. The sample includes one of the most metal poor galaxies known, SBS 0335-052. 

\begin{figure}[t!]
\centering
\includegraphics[width=9cm]{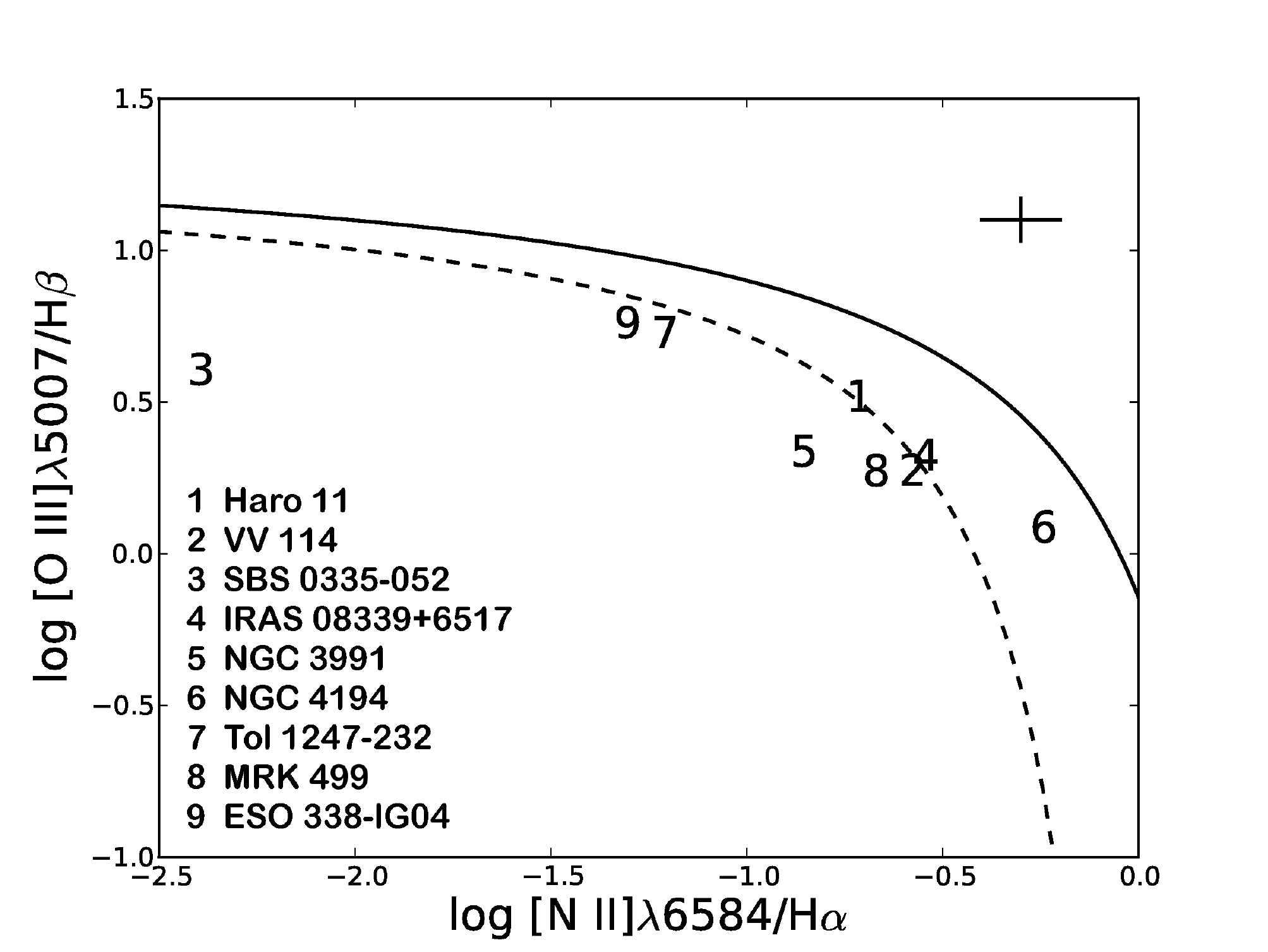}
\caption{The BPT-diagram for  galaxies with available optical spectra. Pure AGNs are found above the solid line \citep{2001ApJ...556..121K}, while pure star forming galaxies are found below the dashed line  \citep{2003MNRAS.346.1055K}.  Typical 1$\sigma$ errors are indicated to the upper right.}
\label{figure:bpt1}
\end{figure}

None of the galaxies have been classified as AGN in the literature. To evaluate possible contributions  from minor or narrow line AGNs to the UV flux, we have  in Fig.~\ref{figure:bpt1} plotted those with available optical spectra in a BPT-diagram \citep{1981PASP...93....5B}.  Pure AGNs are found above the solid line \citep{2001ApJ...556..121K}, while pure star forming galaxies are found below the dashed line  \citep{2003MNRAS.346.1055K}. Between the two lines are  galaxies which consists of either a composite of a star forming and a Seyfert galaxy, or of a star forming galaxy and a LINER. Most galaxies in this sample seem to classify as purely star forming, with the exception of NGC 4194 which seems to be a star forming-LINER composite. The offset of SBS 0335-052 is likely caused by the low metallicity (e.g. \citealt{2006ApJS..167..177D}). Based on this diagnostic, possible AGN effects are not further considered in this paper, and we assume all targets to be mainly ionized by stars.

 The UV luminosities  were derived from the Galaxy Evolution Explorer (GALEX) far-UV (FUV) magnitudes, corrected for Galactic extinction as described in Section~\ref{intabs}. The mean FUV magnitude of the whole sample was $<$M$_{UV}$$>$ = -19.3, as compared to, for example,   the 9 times brighter sample from \citet{2006ApJ...651..688S} with $<$M$_{UV}$$>$= -21.63. The local luminosity function by \citet{2005ApJ...619L..15W} gives M$_{FUV}^{\star}$= -18.04. Most of the galaxies in this sample are brighter than L$_{UV}^{\star}$ in the FUV, and the mean of the sample is  $<$L$_{UV}$/L$_{UV}^{\star}$$>$ = 5.5. The individual FUV luminosities are listed in Table~\ref{table:lyman}.

 While most of the galaxies in this sample have  lower UV luminosities than the typical Lyman Break  Galaxy (LBG),  three of them, Mrk 54, Haro 11, and VV 114,  were classified as Lyman Break Analogues (LBA) in \citet{2011ApJ...730....5H}. These three galaxies do however not have the dominant central object (DCO), that is significant for those galaxies in the Heckman z=0.09-0.23 LBA sample that  show signatures of significant LyC leakage. The DCOs were  identified in \citet{2009ApJ...706..203O}, and seems to be connected to an extreme mechanical feedback from star formation as seen by the outflow velocities ($\Delta$$v_{max}$) derived from interstellar absorption lines. For the leaking DCOs, strong winds were found with $\Delta$$v_{max}$=1500 km s$^{-1}$, while the local LBAs have a factor $\sim$three lower outflow velocities. 

\begin{table*}[t!]
\centering
	\caption{The FUSE observations. (1) The target galaxy, 
	(2) FUSE pointing, the right ascension, (3) FUSE pointing, the declination, (4) The FUSE designation of the data set, 
	(5) The observation date, (6) The exposure time for the 2A detector segment and night only data [s], (7) The MAST category:  82=Irregular Galaxy, 88=Emission-line Galaxy (non-Seyfert), and 89=Starburst Galaxy, (8) LyC=Y: primary science goal was the measurement of the LyC.}
	\begin{tabular}{l l l l l l l l}
	\multicolumn{8}{c}{}\\
	\hline \hline
        Name & RA (J2000) & Dec (J2000) & FUSE ID &  Observation date &Exposure (s)  & Category  & LyC \\  %
        (1) & (2) & (3) & (4) & (5)& (6) & (7) & (8)  \\
     	\hline 
	Haro 11 & $00^h36^m52.7^s$ & $-33^d33^m17.0^s$ & B10901 & 2001-10-12 & 12198 & 88 & Y \\ 
	VV 114 & $01^h07^m46.6^s$ & $-17^d30^m24.0^s$ & C04801 & 2003-07-26 & 4380 &  89 & Y \\
	MRK 0357 & $01^h22^m40.2^s$ & $+23^d10^m10.0^s$ & P19101 & 2000-07-31  & 7117 &  89 & Y \\
	SBS 0335-052 & $03^h37^m44.0^s$ & $-05^d02^m39.0^s$ & A03604 & 2001-09-26 & 13132 &  88 & N \\ 
	Tol 0440-381 & $04^h42^m08.0^s$ & $-38^d01^m10.8^s$ & A05202 & 2000-12-13 & 24189 &  88 & Y \\     
	IRAS 08339+6517 & $08^h38^m23.2^s$ & $+65^d07^m16.0^s$ & 	B00401 & 2001-11-06 & 5299 &  89 & N \\  
	NGC 3991 & $11^h57^m30.8^s$ & $+32^d20^m12.1^s$ & A02302 & 2001-02-14 & 3617 &  82 & N \\ 
	NGC 4194 & $12^h14^m09.7^s$ & $+54^d31^m38.0^s$ & C04803 & 2002-03-30 & 36553 &  89 & N \\  
	Tol 1247-232 & $12^h50^m18.8^s$ & $-23^d33^m57.0^s$ & Q10501 & 2000-05-19/20 & 19316 &  88 & Y \\
	MRK 54 & $12^h56^m55.9^s$ & $+32^d26^m52.0^s$ & A05201 & 2000-02-19 & 15788 &  88 & Y \\  
	MRK 499 & $16^h48^m23.9^s$ & $+48^d42^m33.0^s$ & Q30501 & 2002-07-12 & 12275 &  88 & Y \\  	
	ESO 338-IG04 & $19^h27^m58.0^s$ & $-41^d34^m27.7^s$& A02306 & 2000-05-21 & 2905 &  89 & N  \\
	ESO 185-IG013 & $19^h45^m00.8^s$ & $-54^d15^m01.0^s$ & Z90914 & 2002-08-11 & 7916 &  89 & N  \\  %
    	\hline
	\end{tabular}
	\label{table:ttwo}
\end{table*}

\subsection{The intrinsic absorption}
\label{intabs}
The term \textit{dust extinction}, is a common convention of the combined scattering and absorption effects on light  coming from of a single emitting region. However, when the dust extinction estimates are based on measurements over larger regions (i.e. as from IUE data), different  geometric factors are averaged together and the effect is almost completely dominated by absorption. The term \textit{intrinsic absorption}  \citep{1999ApJ...521...64M} will therefore be adopted  for E(B-V)$_i$ throughout this article.

The intrinsic absorption was calculated from the FUV continuum slope of the galaxies, $\beta$, where f$_{\lambda}$$\propto$$\lambda^{\beta}$. The slope was measured from the IUE SWP spectra, after correction for Galactic extinction using the \citet{1989ApJ...345..245C} extinction law and line of sight values from \citet{1998ApJ...500..525S}. The slope was then fitted with a power law using a second order Legendre polynomial on the log(f$_{\lambda}$)/log($\lambda$) data. A correction factor (-0.16) was applied to the $\beta$ value, as these were derived from the SWP data only (compared to using both SWP and LWP data as  described in \citealt{1999ApJ...521...64M}).  

The internal absorption, E(B-V)$_i$, was then calculated using the \citet{1994ApJ...429..582C} extinction law, but with a modified zero extinction slope. Some of the galaxies in this sample were found to have $\beta$ values bluer than the zero extinction slope derived by Calzetti ($\beta_0$=-1.71), so instead we adopt  $\beta_0$=-2.3 as derived in \citet{1999ApJ...521...64M} in which some of the more massive galaxies had been excluded in the fit. A dependency of $\beta$ with UV luminosity has been observed (e.g. \citealt{2011arXiv1109.0994B}), so this value is likely to be better adapted to most of the galaxies in this sample. In the Meurer paper, a strong correlation between the ratio of far-infrared (FIR) to UV fluxes with spectral slope was observed. Six of the galaxies in our sample were used to derive the IR-to-UV flux ratio/A$_{1600}$ vs $\beta$ relation in Fig.1 of that article, and also the rest of the galaxies here (except VV 114 which lack IUE data) where found to follow the relation closely. 

The E(B-V)$_i$ values are presented in Table~\ref{table:tfour}.

\section{The data}
\subsection{FUSE data}
\begin{figure*}[t!]
\centering
\includegraphics[width=13cm]{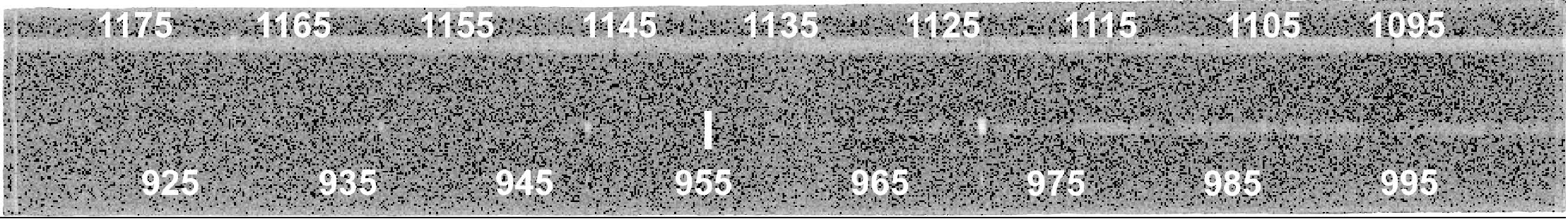}
\caption{The 2A detector segment of Tol 1247-232. The upper spectrum is the LiF channel, and the lower spectrum is the SiC channel. The Lyman limit in the SiC channel is marked with a white vertical line. Geocoronal airglow lines are visible as bright point-like structures. The wavelengths are given in {\AA}. The image is shown in original size with 16384$\times$1024 pixels and on logaritmic scale.}
\label{figure:2d}
\end{figure*}

FUSE is a space telescope with a spectrograph that covers the wavelength range 905-1187 \AA. The FUSE instrument is provided with complementary channels, one based on  silicon carbide (SiC) optics for wavelengths shorter than $\sim$1000 {\AA}, and the other on lithium fluoride (LiF) over-coated aluminum for wavelengths in the range $\sim$1000-1200 {\AA}. The four detectors are divided into two segments  that each image both a SiC and a LiF spectrum, so there are in total eight spectra. Two of these spectra cover the wavelength region where we can measure a possible LyC signal, SiC 1B (905-992 $\AA$) and SiC 2A (916-1005 $\AA$, see Fig.~\ref{figure:2d} for an example). 

The aperture used for all targets was the low resolution aperture (LWRS), with a field of view (FoV)  of 30\arcsec$\times$30\arcsec.  The FUSE observational data of the galaxies have been summarized in Table~\ref{table:ttwo}.

Detailed information about the satellite and its instrument can be found in the \textit{FUSE Archival Instrument
Handbook}\footnote{http://archive.stsci.edu/fuse/ih.html} and \textit{The FUSE Archival Data Handbook}\footnote{http://archive.stsci.edu/fuse/dh.html}. For details about the on-orbit performance of the FUSE instrument, we refer to \citet{2000ApJ...538L...7S}. 

\subsection{Ly$\alpha$ data}
All the galaxies except VV114 have been observed with the IUE, in the low-dispersion mode with the SWP camera. The FoV of the SWP camera is 10\arcsec$\times$20\arcsec, thus less than a quarter of the FUSE FoV. The fully calibrated IUE spectra were downloaded from MAST, and were converted to standard fits-files using the \textit{IRAF}  IUETOOLS package. From these spectra the intrinsic absorption was derived as described in the previous section. Also, the Ly$\alpha$ line equivalent width (EW)  was measured using the \textit{IRAF} task SPLOT with a Gaussian profile for emission lines and a Voight profile for absorption lines. The measured  Ly$\alpha$ quantities are presented in Table~\ref{table:lyman}. 

Four of the galaxies have been observed in high resolution Ly$\alpha$  imaging with the Solar Blind Channel (ACS/SBC) camera onboard the Hubble Space Telescope \citep{2007MNRAS.382.1465H,2009AJ....138..923O}. The FoV of SBC is 34.6\arcsec$\times$30.5\arcsec, hence well matched to the size of LWRS. The Ly$\alpha$ EW derived from the imaging  was found to agree well within the errors to our IUE measurements for three of the galaxies, Haro 11, SBS 0335-052 and ESO 338-IG04. The exception was IRAS 08339+6517, where we measure roughly half of the SBC value (which possibly was overestimated due to the lack of data in the F814W filter, and only one stellar population fitted instead of two).

%

\section{Analysis of the FUSE data}
The FUSE data were partly reduced using the pipeline, CalFUSE v3.2, which is described in \citet{2007PASP..119..527D}. For all our targets the data were obtained using the LWRS aperture in the  time-tag (TTAG) mode. The pipeline corrects for several distortions introduced by the motion of the telescope or by the instrument itself. In the reduction process two-dimensional intermediate data files (IDF) are produced, where the information about every data event is stored. Except for selecting night data only, the data were otherwise screened using the standard set-up. All galaxies in this work are faint in the LyC region, and to improve the S/N all sub-exposures were co-added prior to the background fit as recommended in \citet{2007PASP..119..527D}. At this stage ends the reduction process using CalFUSE, with the co-added IDF files as output product.

The background model of CalFUSE v3.2 is based on three components; one spatially uniform from intrinsic $^{40}$K decay in the micro-channel plates  together with cosmic rays, the other two spatially varying day- and night components from scattered light. The two latter are produced from long observations of a blank sky often months away from the science observations, and it was shown in L11 that the fixed shape of these template background files are not sufficient when working with low S/N FUSE data. Therefore, the one-dimensional files produced by the pipeline are not used in this work. Instead, we use the co-added IDF files and apply a background fit directly to the detector response for these. The procedure is briefly described below, but for details about the new background model, and the analysis on working with low S/N FUSE data, we refer to L11.

\subsection{Extended or point sources}
The spectral height (i.e. the extent of the target in the cross-dispersion direction) of the targets was measured at a position close to the filled aperture airglow Ly$\beta$ emission line in the LiF 1A spectrum, following the method described in Appendix A in L11. The location was chosen since this is close to the minimum of the spectral height for this channel, and the galaxies also have a strong enough continuum here to allow for the comparison. In most cases, the spectral height of the galaxy continuum was found to be roughly half (or slightly less) of that of the airglow emission line. This indicates that most of the galaxies might pass as point sources at non-ionizing wavelengths (rest wavelength $\sim$1000 \AA). However, considering the detections at z$\sim$3 where LyC emitting regions have been observed with large spatial off-sets from the non-ionizing continuum in several cases \citep{2009ApJ...692.1287I,2011ApJ...736...18N}, as well as the possibility of star forming regions residing in the outskirts of our galaxies, we need to view them as possible extended sources in the Lyman continuum. Unfortunately, the FUSE data itself does not offer any spatial information of where the emission comes from.

%


\subsection{The background model}
\label{sec:model}
The two-dimensional IDF files were used to construct the background fit, and the method differs slightly between the two LyC detector segments. 
For the 2A detector segment, a 128$\times$128 pixel sized reconstruction is made of the two-dimensional IDF. A mask is then applied on the regions at the detector edges and where the signal falls. The background is modeled by fitting a smooth surface to the 128$\times$128 masked image using optimal filtering techniques. This is achieved by solving a regularized optimization problem with the solver implemented as an IDL routine: \textit{opt\_filter\_2d.pro}, which is part of the \textit{Reduce} package\footnote{The \textit{Reduce} package and \textit{opt$\_$filter$\_$2d.pro} routine can be found at: \\ http://www.astro.uu.se/$\sim$piskunov/RESEARCH/REDUCE/reduce.html} \citep{2002A&A...385.1095P}. The flux is then calibrated using a linear interpolation between the two CalFUSE effective area files closest in time to the observations of the target galaxy, i.e. the same approach as the standard procedure of CalFUSE. The error in the background fit is propagated through to the final 1D spectrum and the S/N calculations. 

For the 1B detector segment, a slightly different approach is applied. Since the spectrum of interest, the SiC 1B spectrum, falls directly on the steep slope of the detector background, we cannot mask out this region as before. Instead, we use a reconstruction of better resolution, 256$\times$256 pixels, and fit the background with a Gaussian under the spectrum. This is done in an interactive way (see  L11 for details), and five different background images are produced in this manner, with the mean being the final image. The size of the extraction region was found to be an important factor, and was determined for each galaxy individually. We risk missing some of the target signal with this method, but it is essential to exclude the peak of the background slope since this could not be reproduced to a satisfactory level in the background model. 

The one-dimensional spectra were extracted applying these new background models. The spectral resolution was adopted for each target individually to achieve the required S/N needed for the accuracy of the background fit. In the co-addition of the SiC2A and SiC1B spectra,  the SiC1B spectrum which is associated with larger errors is given much lower weight. The final spectrum is therefore dominated by the SiC2A signal. 

\begin{figure*}[ht!]
\centering
\includegraphics[width=9cm]{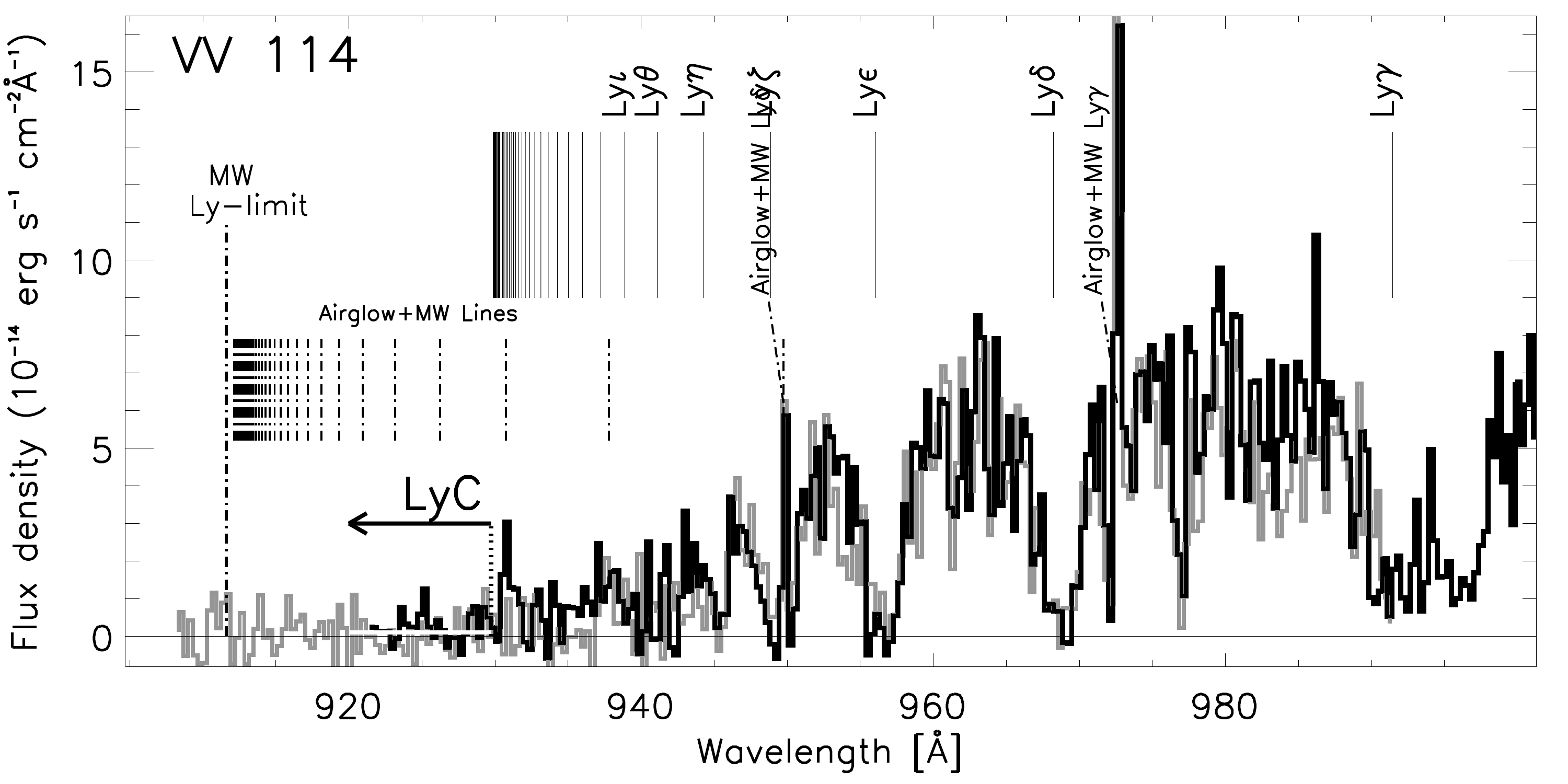}
\includegraphics[width=9cm]{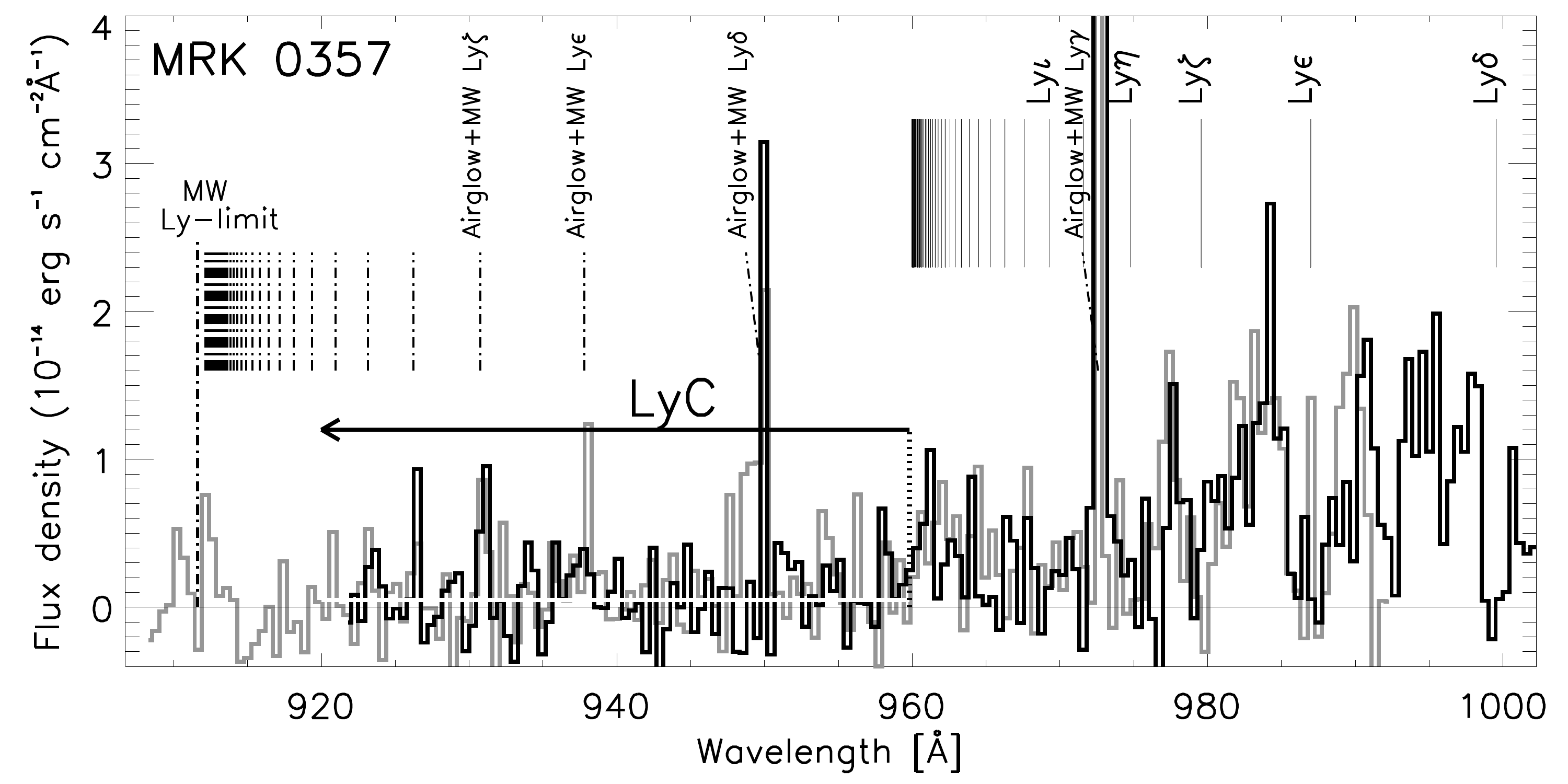}
\includegraphics[width=9cm]{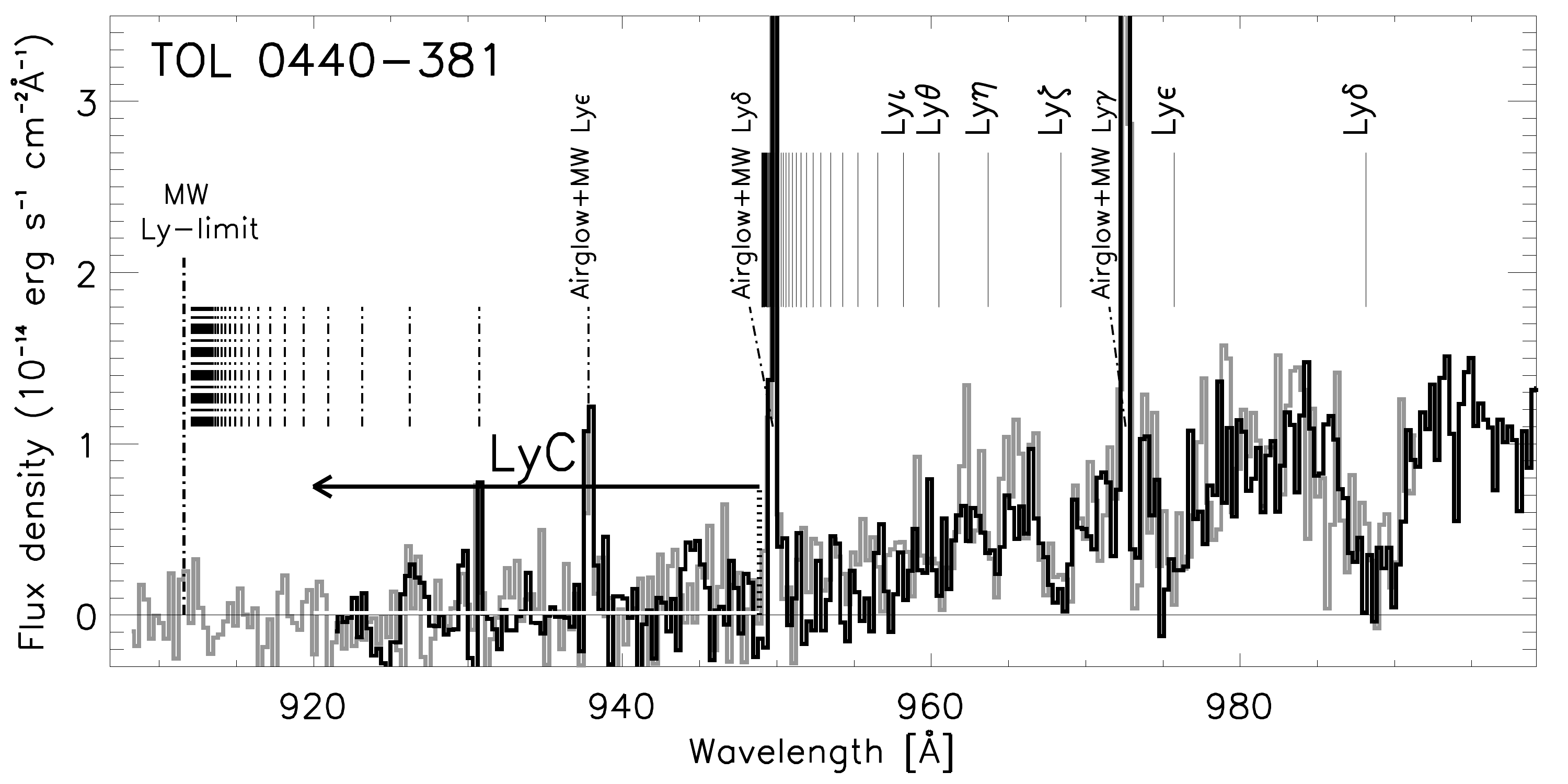}
\includegraphics[width=9cm]{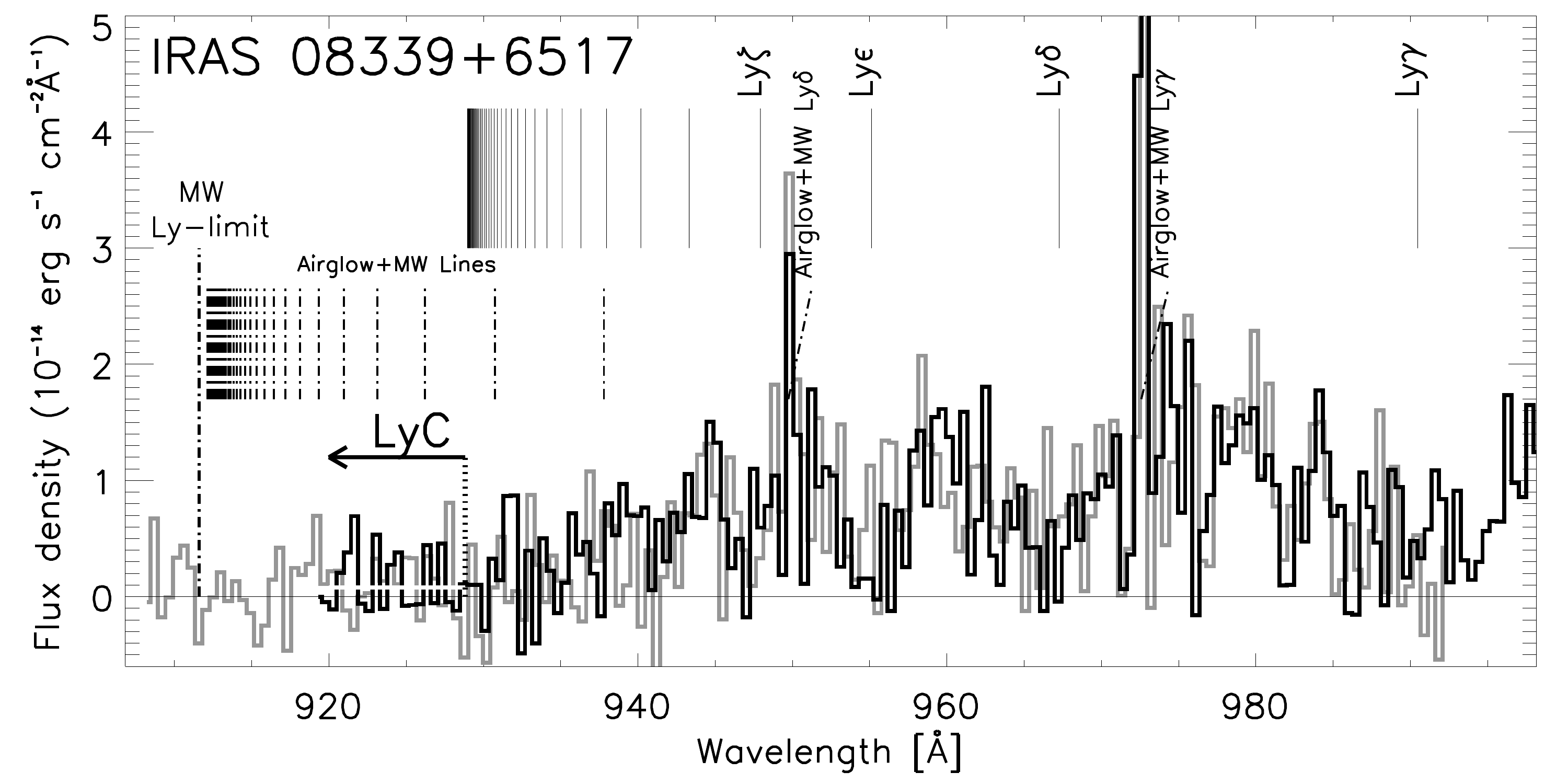}
\includegraphics[width=9cm]{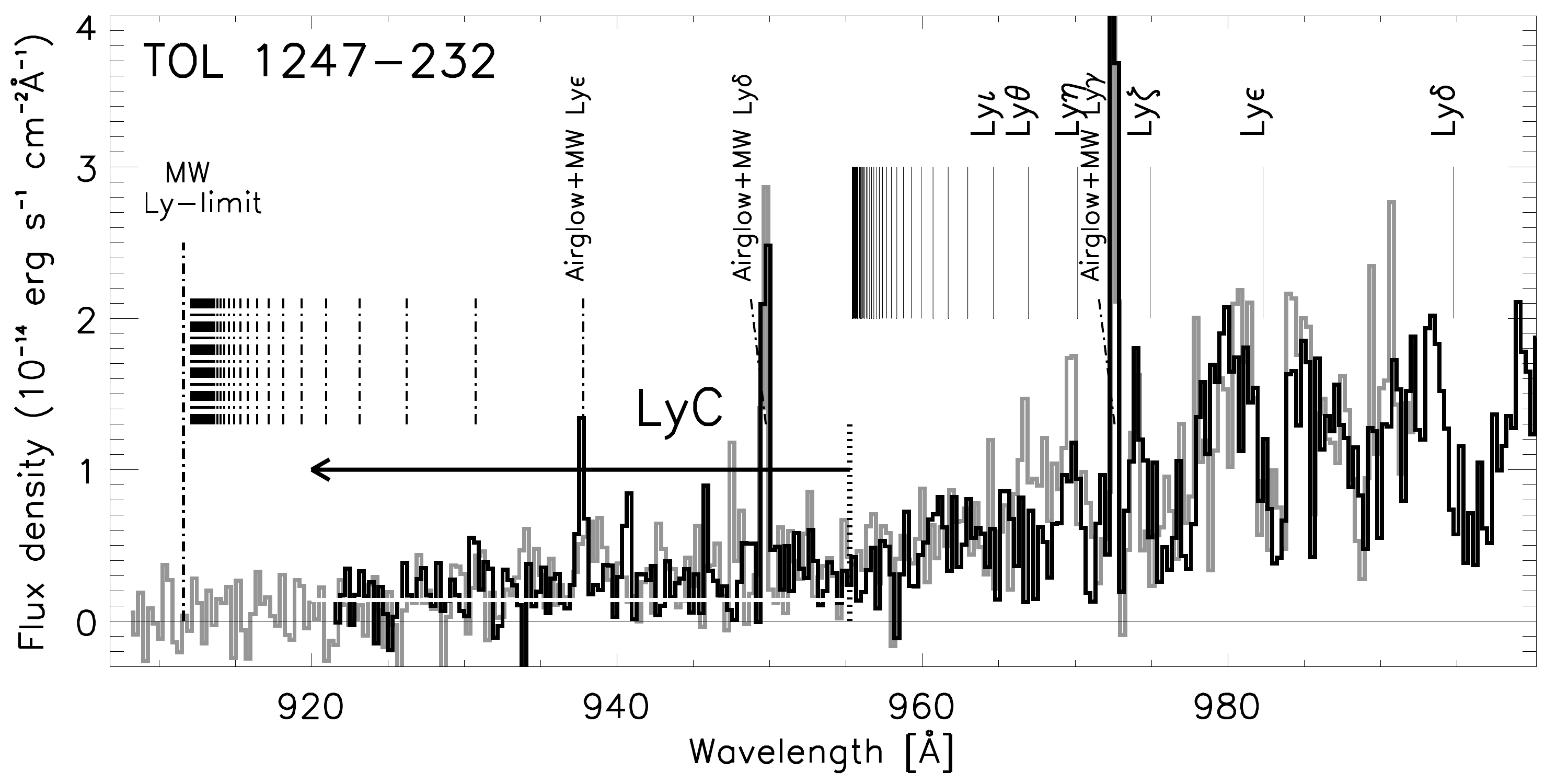}
\includegraphics[width=9cm]{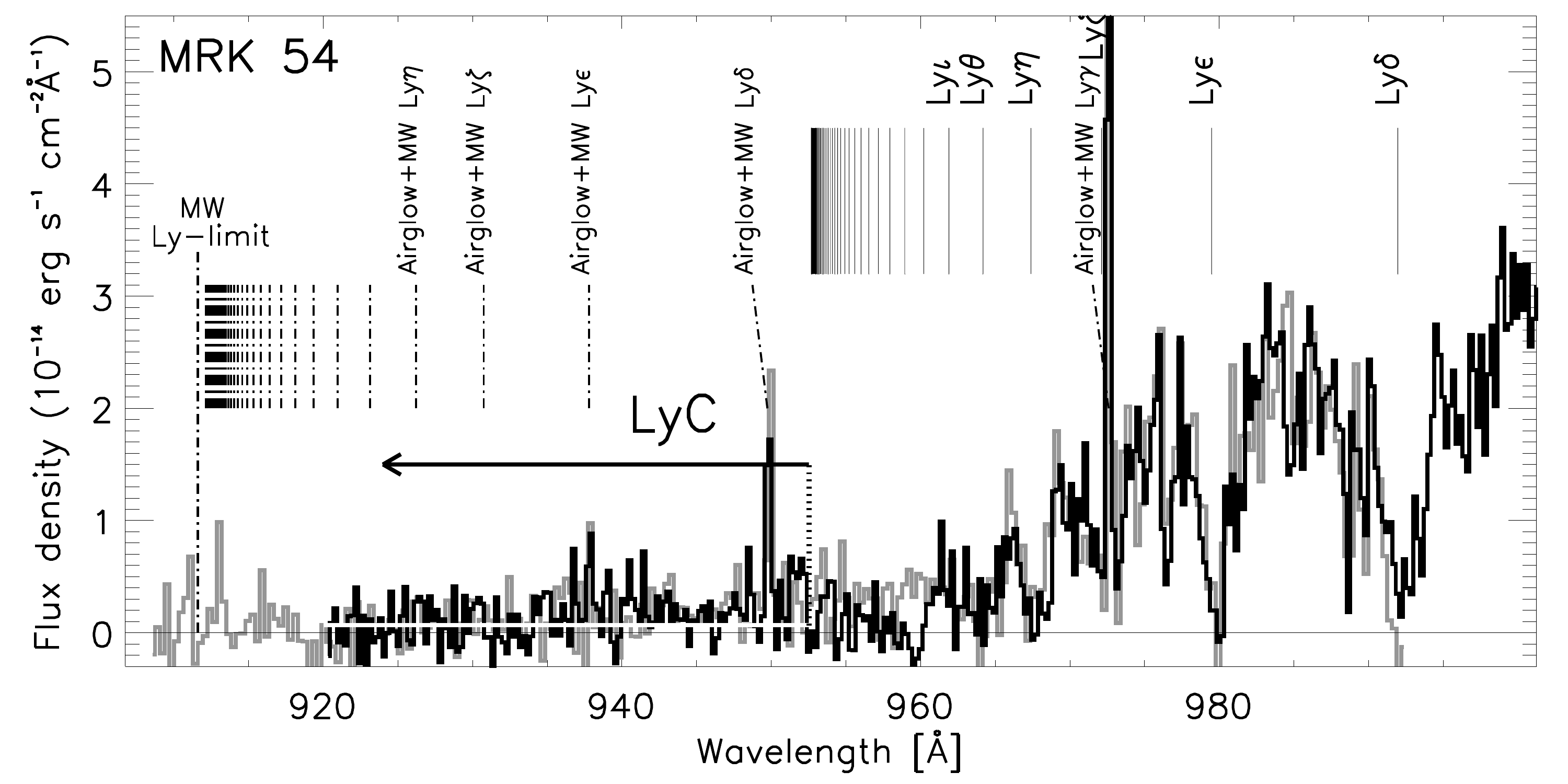}
\includegraphics[width=9cm]{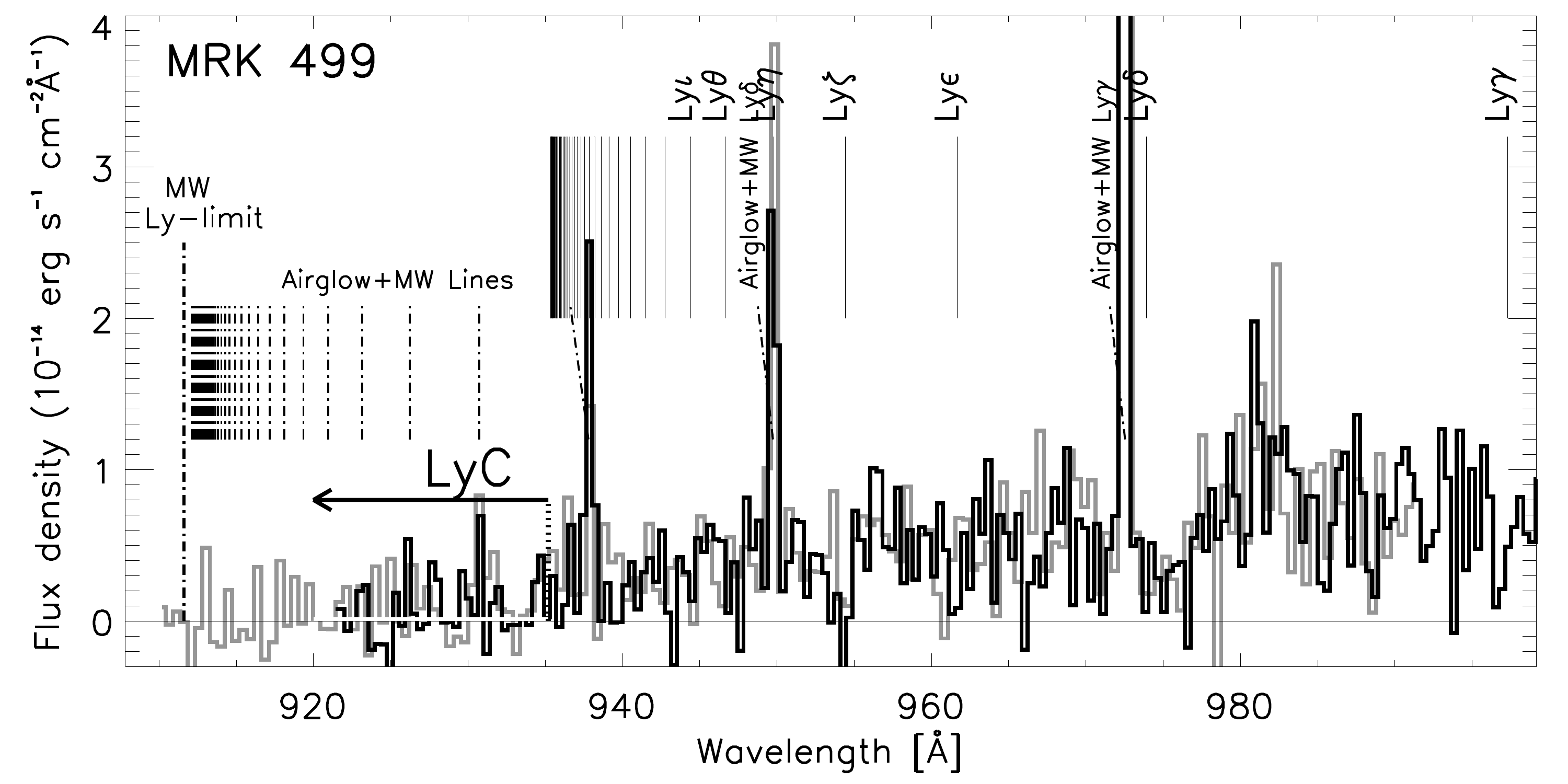}
\includegraphics[width=9cm]{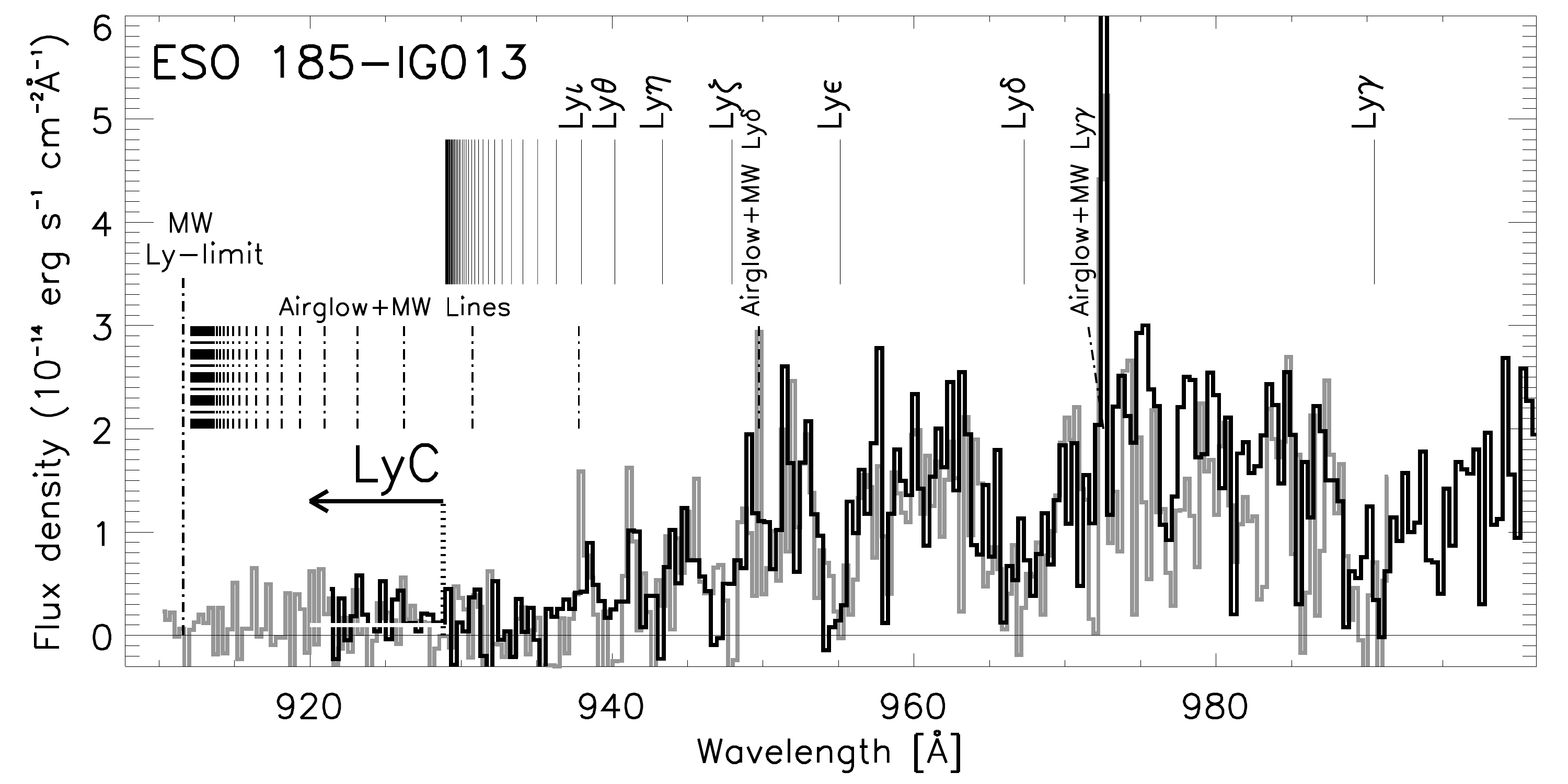}
\caption{The night-only spectra of the eight galaxies for which assessment directly on the Lyman continuum is possible. The SiC 2A data is plotted in black, the SiC 1B data in gray. Marked are also the intrinsic Ly-lines, as well as Milky Way absorption lines with geocoronal airglow emission superposed. The horizontal white line indicates the level of the measured mean of the Lyman continuum.}
\label{figure:leak1}
\end{figure*}

It is also worth to mention that over the mission, the sensitivity  has been found to change differentially between illuminated and un-illuminated regions of the detectors. The background can therefore be somewhat overestimated as compared to the aperture regions using our method. The effect seems to be increasing with time, and would likely be largest for those galaxies observed late in the mission (see Fig.5, L11). We do not try to correct for this effect, but note that it will work towards more conservative measurements of the LyC flux.

\section{Results}

\subsection{LyC measurements below the Lyman limit}
\label{LyC}
The one-dimensional spectra  can be seen in Fig.~\ref{figure:leak1} (excluding Haro 11). The SiC 2A data are plotted in black, and the SiC 1B data in gray. Also, the intrinsic Ly-series lines are marked, as well as the Milky Way Lyman absorption lines with geocoronal emission superposed. The LyC flux was measured on the airglow free regions below the Lyman limit, and the mean flux density level is indicated by a white horizontal line. As mentioned, the final LyC fluxes were calculated from a co-added spectrum using weights based on the error in the background fit. Due to the more reliable background fit in the SiC 2A spectrum, this channel dominates the final signal in all cases. The following flux densities are corrected for Galactic extinction as described in Section~\ref{fesc}, and the errors include both the systematic error in the background fit and the statistics in the 1D spectrum. 

A clear excess was found in the LyC only for one of the galaxies, Tol 1247-232, for which a flux density of 
$f_{900}$=4.4 $\times$ 10$^{-15}$ erg s$^{-1}$ cm$^{-2}$ \AA$^{-1}$ (S/N=5.2) was measured. For a comparison, the two-dimensional SiC 2A spectrum can be seen in Fig.~\ref{figure:2d}, displaying a weak signal in the continuum below the Lyman limit.

Also Mrk 54 displays a weak LyC signal excess, $f_{900}$=1.0 $\times$ 10$^{-15}$ erg s$^{-1}$ cm$^{-2}$ \AA$^{-1}$, although less significant with S/N=1.4. 

For the other galaxies only upper limits to the LyC flux density were derived (95 $\%$ confidence limit): VV114: $f_{900}$ $<$ 2.1 $\times$ 10$^{-15}$, MRK 0357: $f_{900}$ $<$ 2.5 $\times$ 10$^{-15}$, Tol 0440-381: $f_{900}$ $<$ 0.2 $\times$ 10$^{-15}$, IRAS 08338+6517: $f_{900}$ $<$ 4.9 $\times$ 10$^{-15}$, MRK 499: $f_{900}$ $<$ 0.2 $\times$ 10$^{-15}$, 
ESO 185-IG013: $f_{900}$ $<$ 3.6 $\times$ 10$^{-15}$ erg s$^{-1}$ cm$^{-2}$ \AA$^{-1}$.

The flux densities are listed in Table~\ref{table:lyman}.

\subsection{The escape fraction,  $f_{esc}$ }
\label{fesc}
The absolute escape fraction of ionizing photons,  $f_{esc}$, is the ratio of the observed LyC flux density  to the intrinsic stellar LyC flux density. Here, for our sample of local galaxies, we evaluate the escape fraction at rest frame 900 {\AA}. The intrinsic ionizing flux density produced by the stars  at 900 {\AA} ($f_{900,\mathrm{int}}$) can be estimated by the integrated H$\alpha$ flux corrected for dust absorption, so in order to calculate this ratio we need to have detailed information about the H$\alpha$ flux as well as the dust absorption within the region covered by the FUSE LWRS aperture. Such detailed information is not available for all our targets, and we will instead use the higher redshift approach of measuring the relative escape fraction ($f_{esc,rel}$). This will also have the advantage of making it easier to compare this work to other results.

Since the H$\alpha$ line is no longer available as a diagnostics for high-z galaxies,  the 1500 {\AA} flux density is usually used instead. Here,  $f_{esc,rel}$ represents the ratio between the intrinsic Lyman break amplitude, ($f_{1500}/f_{900})_\mathrm{int}$, and the observed ratio, ($f_{1500}/f_{900})_\mathrm{obs}$. Assuming a picket-fence model, we can compare the absolute and relative escape fractions by correcting for internal absorption at 1500 {\AA} ($A_{1500,i}$):


\begin{eqnarray}
f_\mathrm{esc} = f_\mathrm{esc,rel}  \times 10^{-0.4 \times A_{1500,i}} = \frac{(f_{1500}/f_{900})_\mathrm{int}}{(f_{1500}/f_{900})_\mathrm{obs}}  \times 10^{-0.4 \times A_{1500,i}}.
\label{eqfesc}
\end{eqnarray}

 where $A_{1500,i}$ was calculated from E(B-V)$_i$, using the \citet{1997AJ....113..162C} extinction law. The intrinsic ratio, ($f_{1500}/f_{900})_\mathrm{int}$, is usually evaluated by models and is further discussed in Section~\ref{intratio}. Since we are working with local galaxies, we do not make any corrections for the IGM optical depth. 
 
 The observed $f_{1500}$ was measured on line-free regions in the rest-frame 1500 {\AA} continuum of the IUE spectra. To derive ($f_{1500}/f_{900})_\mathrm{obs}$, both fluxes were corrected for Galactic extinction using the \citet{1989ApJ...345..245C} law down to 1250 {\AA}, and from 1250 to 900 {\AA}  the \citet{1990ARA&A..28...37M} extinction law, with line of sight values from \citet{1998ApJ...500..525S}. 
 
As mentioned in Section 3.2, the FoV of IUE and FUSE are not identical. While most of the FUV light should be included in both cameras for the smaller galaxies, the uncertainty in ($f_{1500}/f_{900})_\mathrm{obs}$ will be higher for the larger galaxies. Off-center LyC leaking regions could also increase the uncertainty, but as we have no spatial information we can not make any correction for the morphology of the galaxies. All further calculations are made under the assumption that the majority of the FUV emission comes from within both  the FUSE and the IUE FoV.

\subsubsection{The intrinsic ratio, ($f_{1500}/f_{900})_\mathrm{int}$}
\label{intratio}
The ($f_{1500}/f_{900})_\mathrm{int}$ parameter is model dependent, and for the young burst ages that most of the galaxies in this sample show signs of, the ratio takes values  between 1-2  (in f$_{\lambda}$) using either of the \citet{2001A&A...375..814Z} or Starburst99 \citep{1999ApJS..123....3L} models. Young stars dominate the SED in the FUV, and  the intrinsic ratio has been found to be strongly dependent on the star formation history within the last $<$ 10 Myr   (e.g. \citealt{2010ApJ...723..241S}). It was also argued, in the same paper, that SED fitting alone is not sufficient to derive ($f_{1500}/f_{900})_\mathrm{int}$, but that additional data like H$\alpha$ measurements are necessary. This approach was attempted for Haro 11 in \cite{2006A&A...448..513B}, but it was found that the H$\alpha$ constraint was not strong enough to significantly improve the result.

In L11,  an escape fraction for Haro 11 of $f_{esc}$=3.3$\pm$0.7 $\%$ was derived. The calculation was based on $f_{900,\mathrm{int}}$ computed from the bin-wise SED modeling by \citet{2007MNRAS.382.1465H}, where the production of LyC photons was reconstructed over the whole galaxy. The modeling was performed using a multiple filter setup, ranging from the FUV to near-infrared (NIR), including the H$\alpha$ line and with a good sampling of the Balmer break. Thus, ($f_{1500}/f_{900})_\mathrm{int}$ is  well determined for Haro 11 at each modeled spaxel, and we should in principle be able to derive the mean value for the whole galaxy  from Eq.~\ref{eqfesc}. This would be the best constrained modeled   ($f_{1500}/f_{900})_\mathrm{int}$ to date.  The  value  derived in \citet{2007MNRAS.382.1465H}, $f_{900,\mathrm{int}}$= 12.3 $\times$ 10$^{-14}$ erg s$^{-1}$ cm$^{-2}$ \AA$^{-1}$,  was stated without error. Accounting for flat-fielding errors and the absolute calibration of HST, the error in the 1500 {\AA} flux will be roughly 3 percent. However, the extrapolation of the model spectra from 1500 to 900 {\AA} may introduce a substantially larger error component that results from statistical uncertainty in the recovered stellar ages. Should each resolution element contribute equally to the total flux, this uncertainty would be expected to be very large, but fortunately the huge majority of FUV light is output from a small fraction of pixels -- those that are both bright and very blue. This immediately limits the range of permitted ages for the resolution elements that dominate the FUV (see also \citealt{2010MNRAS.407..870A}), and over these limited age ranges we estimate a dispersion of around 30 percent on the total $f_{900,\mathrm{int}}$ flux. This estimate is based upon statistical considerations only, and accounts for no systematic component that may result from the severely limited set of observational constraints placed upon the ionizing output of the most massive stars. If we translate the uncertainty in $f_{900,\mathrm{int}}$ into the escape fraction for Haro 11, we arrive at $f_{esc}$=3.3$^{+1.2}_{-1.0}$ $\%$.

The observed ratio in Haro 11 is ($f_{1500}/f_{900})_\mathrm{obs}$=9.3, which gives an intrinsic ratio of ($f_{1500}/f_{900})_\mathrm{int}$ $\approx$ 1.5$^{+0.6}_{-0.5}$ (or 4.2$^{+1.7}_{-1.4}$ in $f_{\nu}$), and the error in  ($f_{1500}/f_{900})_\mathrm{int}$ thus includes (and are dominated by) the errors in $f_{900,\mathrm{obs}}$ and $f_{900,\mathrm{int}}$. The relative escape fraction of Haro 11 is  $f_{esc,rel}$ = 16.6$^{+7.4}_{-6.5}$ $\%$. The round-off to ($f_{1500}/f_{900})_\mathrm{int}$ $\approx$ 1.5 gives   $f_{esc}$=3.2$^{+1.2}_{-1.0}$ $\%$, which is the value used for Haro 11 throughout the rest of the analysis in this work.

Using the same value of the intrinsic ratio for Tol 1247-232 could be well motivated, since there are several indications of a very young starburst also for this galaxy (see Section~\ref{individ}), similar to that of Haro 11.  The observed ratio of Tol 1247-232 was  ($f_{1500}/f_{900})_\mathrm{obs}$ = 8.6, which gives a relative escape fraction of  $f_{esc,rel}$ = 17.4$^{+7.7}_{-6.7}$ $\%$.  

\begin{figure*}[t!]
\centering
\includegraphics[width=9cm]{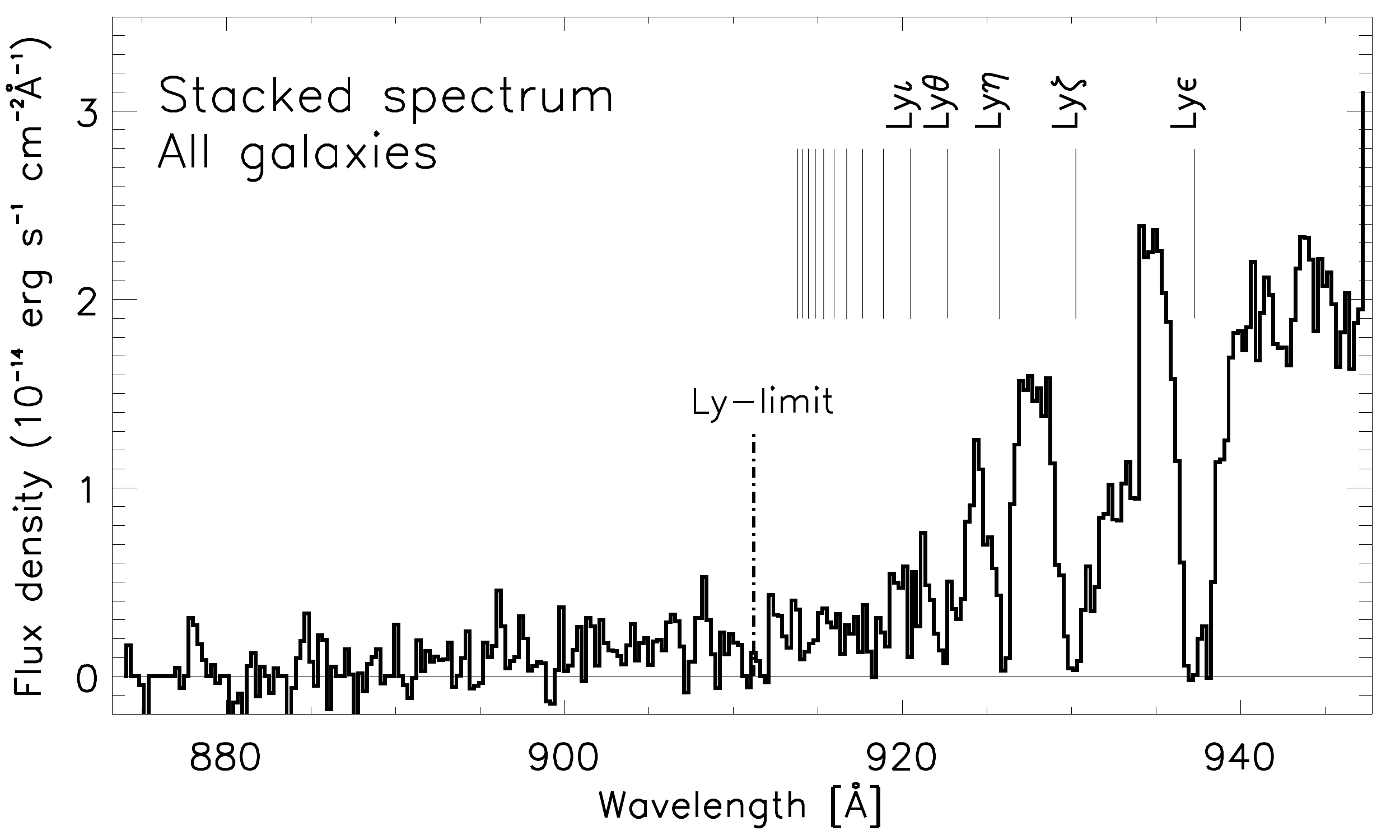}
\includegraphics[width=9cm]{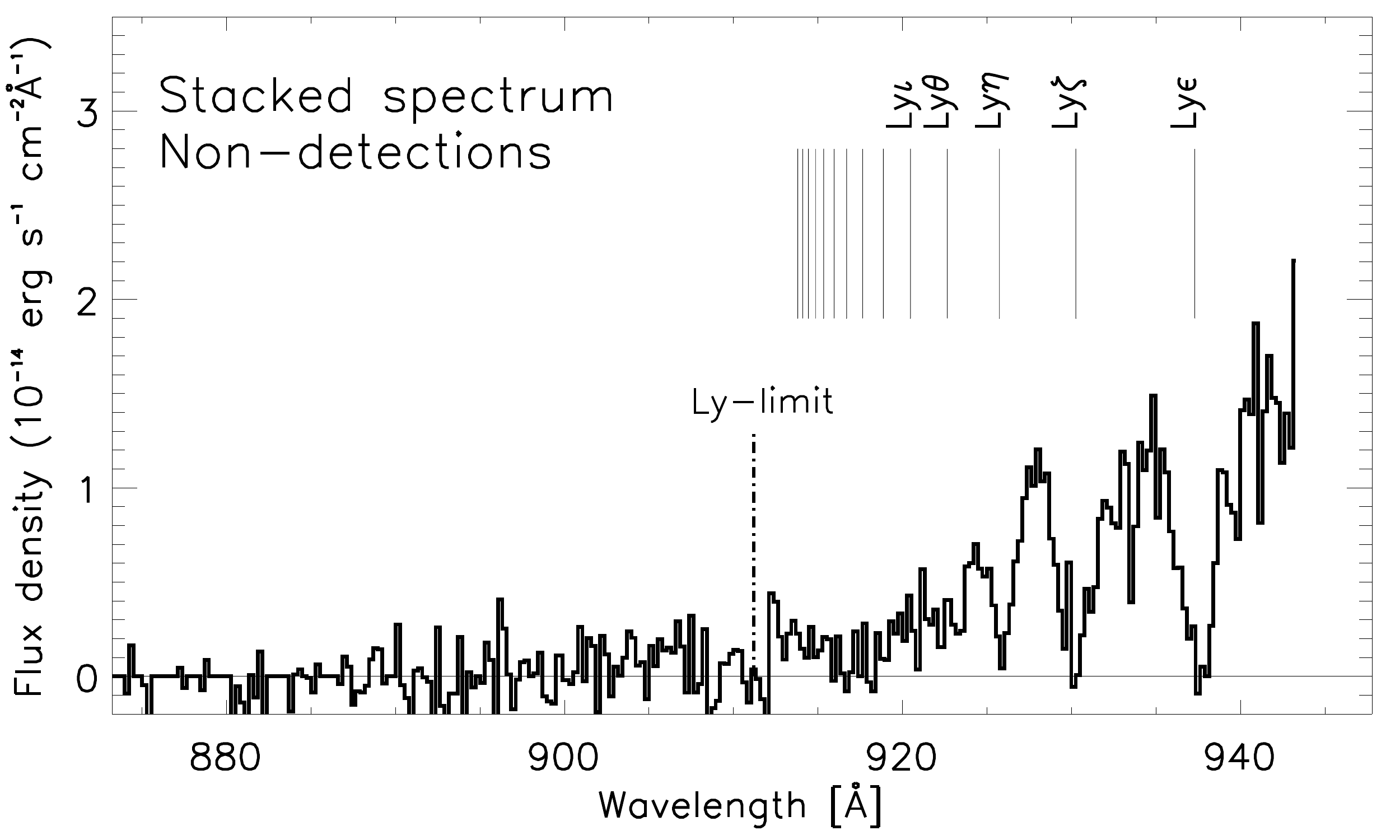}
\caption{ \textbf{Left:} The stacked spectrum of all nine galaxies in the LyC sample,  \textbf{Right:} The stacked spectrum of the seven galaxies non-detected in the LyC. The scale is 0.27 {\AA} per pixel, and the spectra have been blue-shifted into rest-frame. }
\label{figure:stack}
\end{figure*}


\subsubsection{Results on the absolute escape fraction}
 The absolute escape fraction  for Tol 1247-232 using ($f_{1500}/f_{900})_\mathrm{int}$=1.5$^{+0.6}_{-0.5}$  was found to be $f_{esc}$=2.4$^{+0.9}_{-0.8}$ $\%$.  In the derivation of this value, the 1500 {\AA} flux was corrected for internal dust absorption as derived by the IUE UV slope (E(B-V)$_i$=0.21). We note that \citet{2002A&A...393...33B}  found that the reddening based on the far-UV spectral slope in the FUSE spectrum must be almost zero, inconsistent with the IUE UV slope. If the absorption actually is lower, our value of 2.4$^{+0.9}_{-0.8}$ $\%$ is a lower limit to the escape fraction, and in the extreme case of no dust absorption,  $f_{esc}$ = 17.4$^{+7.7}_{-6.7}$ $\%$. The origin of this discrepancy remains unknown but will be addressed in a future paper based on new HST data. 

For MRK 54 an escape fraction of  $f_{esc}$=1.2$^{+1.2}_{-1.1}$ $\%$ was found, which might be interpreted as a sign of weak leakage.

The upper limits of $f_{esc}$ for the rest of the sample were calculated conservatively, adopting the upper limit on the intrinsic ratio $(f_{1500}/f_{900})_\mathrm{int}$=2.1. The 95$\%$ confidence upper limits on $f_{esc}$ are listed together with the other results  in Table~\ref{table:lyman}.

 %

\subsection{The stacked sample}
\label{stack}

At higher redshifts there have been several articles published where the individual spectra have been co-added into a single spectrum, which is of great interest for determining global properties at different epochs. At z $\approx$ 0.7 \citet{2010ApJ...720..465B} derived an upper limit of $f_{esc}$ $<$1 $\%$ by stacking 18 LBAs. At higher redshifts, z $\approx$ 3, there have been two previous articles with this approach. The observed ratio ($f_{1500}/f_{900})_\mathrm{obs}$=17.7$\pm3.8$ in \citet{2001ApJ...546..665S}, translates into $f_{esc}$=23.5$\pm5.0$ $\%$, and the stacked sample in \citet{2006ApJ...651..688S} with ($f_{1500}/f_{900})_\mathrm{obs}$=58$\pm35$, gives $f_{esc}$=6.7$\pm4.0$ $\%$ (for E(B-V)$_i$=0.1, ($f_{1500}/f_{900})_\mathrm{int}$=4 (f$_\nu$), and $\tau_{IGM}$=0.99 and 0.92, respectively). However, both the Shapley and Steidel results have been questioned lately with the refinement of methods to detect foreground interlopers. The LyC signal in the stack of \citet{2006ApJ...651..688S} is dominated by  the two single detections D3 and C49, none of which have been confirmed in follow-up observations (\citealt{2009ApJ...692.1287I, 2011ApJ...736...18N,2012arXiv1210.2393N}).


Since we have nine local galaxies with sufficiently high redshifts (all galaxies with filled rows under f$_{esc,LyC}$ in Table~\ref{table:lyman}), and hence baselines to measure the LyC on, for the first time an attempt to derive the escape fraction in the stacked spectrum  of local galaxies is made. The sample is not selected on the basis of containing likely high-redshift analogues; we simply take what we have at hand. Still, they were all originally selected to be observed by FUSE for their star forming qualities. Some of the galaxies are even known to be among the most powerful local starbursts, and can be considered as scaled down versions of  z $\sim$ 3 LBGs. In many regards these galaxies have properties similar to the LBGs, such as the UV surface brightnesses and star formation rates per unit area (e.g. \citealt{1999ApJ...521...64M,2008ApJ...677...37O}).  There are other parameters where they differ, but while the sample is 9 times fainter in the UV than  the galaxies from \citet{2006ApJ...651..688S}, they appear  more similar to the fainter LAEs in \citet{2011ApJ...736...18N,2012arXiv1210.2393N}. Our sample also  have $<$E(B-V)$_i$$>$=0.19, which is higher than the typical z$\sim$3 galaxy (e.g. in \citealt{2006ApJ...651..688S} $<$E(B-V)$_i$$>$=0.11). The mean redshift of the stacked sample is $<$z$>$=0.03.

The stack was produced as follows. First all airglow regions were given zero weight before the individual spectra were blue-shifted into their rest frame. Only data with observed wavelengths above 920 {\AA} were used. The spectra were then co-added, using the  S/N in the stellar continuum above the Lyman limit at rest-frame $\sim$945 {\AA} as weight.  Two spectra were produced, one with all nine galaxies in the LyC sample, and one with the seven galaxies non-detected in the LyC. The stacked spectra are displayed  in Fig.~\ref{figure:stack}.  

\begin{figure*}[ht!]
\centering
\includegraphics[width=18cm]{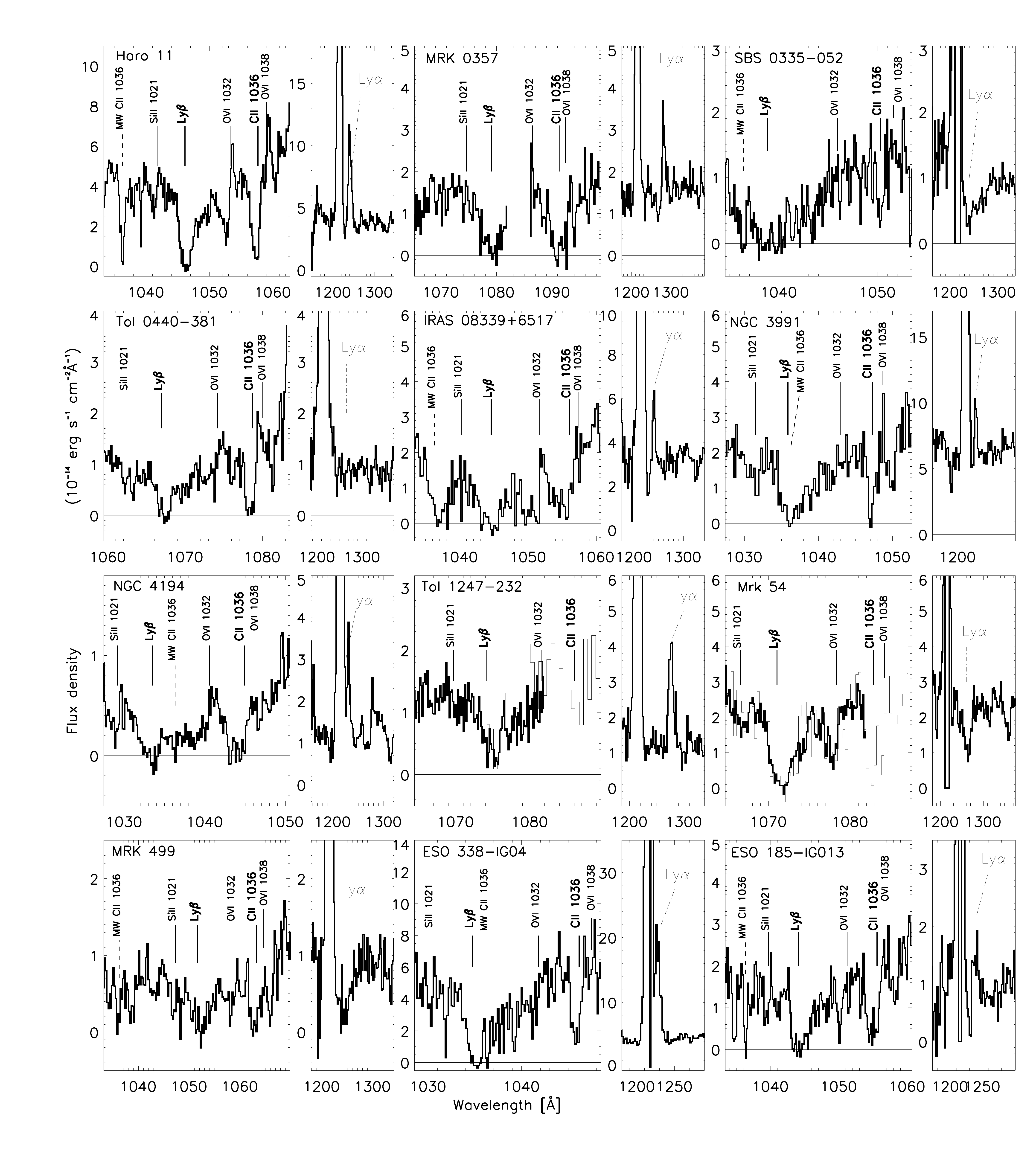}
\caption{Some important spectral lines of the twelve galaxies observed with both FUSE and IUE. \textbf{Left} panels are the FUSE data around Ly$\beta$ and \ion{C}{ii} $\lambda$1036 \AA. Also, when possible, the Milky Way \ion{C}{ii} $\lambda$1036 {\AA} absorption feature is displayed, which shows the zero flux level. Most FUSE data are from the LiF 1A channel, but for the highest redshift targets LiF 2A (MRK 0357) or SiC 2B data are over-plotted (Tol 1247-232, MRK 54). \textbf{Right} panels are the IUE data around the Ly$\alpha$ with the strong geocoronal Ly$\alpha$ emission visible out of focus to the left. }
\label{figure:multispec}
\end{figure*}

\begin{table*}[htb!]
	\centering
	\caption{(1) The target galaxy, (2) The Lyman continuum flux density, after correction for Galactic extinction. Upper limits are 95 $\%$ confidence limits [10$^{-15}$ erg s$^{-1}$ cm$^{-2}$ \AA$^{-1}$], (3) The absolute escape fraction measured in the LyC [$\%$], (4) The relative escape fraction at rest wavelength 1036 {\AA} calculated from the residual flux in \ion{C}{ii} $\lambda$1036 {\AA}  [$\%$], (5) The absolute escape fraction at rest wavelength 900 {\AA} calculated from the residual flux in \ion{C}{ii} $\lambda$1036 {\AA}  [$\%$], (6) The Ly$\alpha$ equivalent width [\AA], (7) The log of  \ion{H}{I} mass [\sma], (8) The log of stellar mass (see Section ~\ref{corr} for details) [\sma], (9) The star formation rate, SFR=SFR(IR)+SFR(UV) \citep{1998ARA&A..36..189K} [\sma yr$^{-1}$], (10) The  SFR normalized by stellar mass (specific SFR, SSFR) [10$^{-9}$ yr$^{-1}$], (11) The GALEX FUV luminosity compared to the local L$_{UV}^{\star}$ by \citet{2005ApJ...619L..15W}.} 
	\begin{tabular}{l l l l l l l l l l l}
	\multicolumn{11}{c}{}\\
	\hline \hline
        Name & $f_{900,obs}$ & $f_{esc,LyC}$ & $f_{esc,rel,CII}$  &$f_{esc,CII}$  &   EW(Ly$\alpha$) & log \mass$_{\ion{H}{i}}$ &  log \mst & SFR  & SSFR & L$_{UV}$/ \\ 
         & 10$^{-15}$ cgs & $\%$ & $\%$ & $\%$ &  [\AA] & [\sma] & [\sma] & [\sma yr$^{-1}$] & [10$^{-9}$ yr$^{-1}$]   & L$_{UV}^{\star}$\\
        (1) &  (2) & (3) & (4) & (5)& (6) & (7) & (8)  &   (9)  & (10) & (11) \\
        	\hline 
	\smallskip
	Haro 11 & 4.0$\pm$0.9 & 3.2$^{+1.2}_{-1.0}$ & 8.1$\pm$1.2 & 2.4$\pm$0.4 & 18.9 & $<$8.7 & 10.2 & 23.8 &   1.5  & 3.5 \\ 
	\smallskip
	VV 114 & $<$2.1 & $<$ 0.4 & 3.9$\pm$0.9 & $<$ 0.2 & -- & 9.7 & 10.6 & 57.3 &  1.1 & 6.1 \\ 
	\smallskip
	MRK 0357 & $<$2.5 &  $<$ 2.3 & 4.5$\pm$1.1 &  $<$ 0.9  & 8.9 & 10.0 &  10.6 & 34.6 &  0.9 & 16.1 \\ 
	\smallskip
	SBS 0335-052 & -- & -- & 16.8$\pm$4.9$^a$ & 16.8$\pm$4.9 & -31.7  & 8.9 &  8.3 & 0.6 &  2.8 & 0.4 \\ 
	\smallskip
	Tol 0440-381 &$<$0.2 & $<$ 0.3 & 5.3$\pm$1.7 &  $<$ 0.6  & 0 & -- &  10.0 & 7.7 & 0.8 & 3.9 \\ 
	\smallskip
	IRAS 08339+6517 & $<$4.9 & $<$ 6.4 & 3.8$\pm$1.3 & 2.8$\pm$1.0 & 14.9 &  9.9 &  10.6 & 20.5 &  0.5 & 5.5 \\ 
	\smallskip
	NGC 3991 & --  & -- & 0.0 &  0.0 & 3.2 & 10.6 & 10.0 & 4.1 &  0.4 & 2.4 \\ 
	\smallskip
	NGC 4194 & -- & -- & $<$4.6 & $<$ 0.1 & 9.2 & 9.4 &  10.3 & 12.1 & 0.6 & 0.5 \\ 
	\smallskip
	Tol 1247-232 & 4.4$\pm$0.8 & 2.4$^{+0.9}_{-0.8}$ & -- & --  &   28.6 & -- & 9.7 & 27.1 &  5.4 & 12.0 \\ 
	\smallskip
	MRK 54  & 1.0$\pm$0.7 & 1.2$^{+1.2}_{-1.0}$ &  $<$5.9 &  $<$2.9  & -14.2 & 10.2 &  10.5 & 28.0 &  0.9 & 16.6 \\ 
	\smallskip
	MRK 499 & $<$0.2 & $<$ 0.5 & $<$ 2.5&  $<$ 0.2 & -22.5 & $<$10.4 &  10.2 &  - & - & 1.6 \\ 
	\smallskip
	ESO 338-IG04 & --& -- & 18.1$\pm$1.5 & 16.2$\pm$1.3 & 28.3 & 9.1 &  9.1 & 2.5 &  1.7&  1.6 \\ 
	\smallskip
	ESO 185-IG013 & $<$3.6 &  $<$ 16.0 & 12.4$\pm$1.8 & 6.0$\pm$0.9  & 5.0 & 9.5 &  9.6 & 2.9 &  0.7 & 1.7\\ 
	\hline
	\end{tabular}
	\\
	\scriptsize{$^a$ Measured on the best S/N spectrum, LiF1A. The LiF2B spectrum is disagreeing, and gives zero flux at the line core. }
	\label{table:lyman}
\end{table*}

 For the full LyC sample, an excess in the LyC between 890-910 {\AA} was measured, with $f_{900}$ = 2.3 $\times$ 10$^{-15}$ erg s$^{-1}$ cm$^{-2}$ \AA$^{-1}$ (S/N=3.4). For the stack of the non-detected galaxies no obvious excess was measured ($f_{900}$ = 2.8 $\times$ 10$^{-16}$ erg s$^{-1}$ cm$^{-2}$ \AA$^{-1}$, S/N=0.6). As in \citet{2006ApJ...651..688S}, we note that the two single LyC detected galaxies (here Haro 11 and Tol 1247-232), are the main contributors to the ionizing flux in the stacked spectrum. 

To evaluate the escape fraction, the mean flux density at rest wavelength 1500 {\AA} was derived from the IUE spectra using the same weighting as for the FUSE data (excluding VV 114 which lack IUE data), with $<$$f_{1500}$$>$=4.1 $\times$ 10$^{-14}$ erg s$^{-1}$ cm$^{-2}$ \AA$^{-1}$. The mean intrinsic  absorption for the sample is $<$E(B-V)$_i$$>$=0.19, while the mean Galactic extinction is $<$E(B-V)$_G$$>$=0.042.  The observed ratio was  ($f_{1500}/f_{900})_\mathrm{obs}$ = 17.8, resulting in a relative escape fraction of $f_{esc,rel}$ = 8.4$^{+3.7}_{-3.2}$ $\%$.

The absolute escape fraction for the stacked full LyC sample was found  to be $f_{esc}$ = 1.4$^{+0.6}_{-0.5}$ $\%$. 

\subsection{LyC estimates from \ion{C}{ii} $\lambda$1036 \AA}
\label{Cii}

This method was introduced in \citet{2001ApJ...558...56H}, and is based on the expectation of a residual flux in the low ionization interstellar absorption line  \ion{C}{ii} $\lambda$1036 {\AA} if the galaxy is leaking ionizing photons. The method invokes a simple picket-fence model where the ionizing photons are escaping through holes in the gas. The method was slightly elaborated  in L11 where the fact that we are likely looking at two different components were entered into the calculations in more detail.  While one component is affected by gas and dust attenuation (at all wavelengths),  the component that is escaping through holes is not affected by gas or dust at all. Further, we are probably looking at two different stellar populations. Through the holes the light is likely dominated by the very young ionizing population, while through the gas we see a more mixed population (including ionizing stars). These effects can play a significant role when we extrapolate the escape fraction from rest wavelength 1036 {\AA} down to 900 {\AA}, which is illustrated for Haro 11 in Fig.11 of  L11. If we assume that the young population we see through the holes has a mean age of 3 Myr and the non-leaking mixed population a mean age of 20 Myr, then the extrapolation can be described by the relation:

\begin{eqnarray}
f_{LyC,CII} = 1.65 \times f_{esc,rel,CII} \times 10^{-0.4 \times 11.1 \times E(B-V)_i},
\label{eq:cii}
\end{eqnarray}

where f$_{esc,rel,CII}$ is the ratio of the residual flux in the line over the continuum flux density. In the derivation of Eq.~\ref{eq:cii}, the \citet{2001A&A...375..814Z} SED model was used with a Salpeter IMF, constant SFR, and 20 $\%$ solar metallicity.

The escape fraction, $f_{esc,CII}$, was estimated from the residual flux in the \ion{C}{ii} $\lambda$1036 {\AA} line as compared to the continuum flux (ratio=f$_{esc,rel,CII}$), and is listed in Table~\ref{table:lyman}. This relation is based on the assumption that the line is purely interstellar. However, in  \citet{2001ApJ...558...56H} they make a comparison with the model starburst spectra by \citet{1997ApJ...489..601G} and find that the stellar photospheric line can depress the continuum by a factor of 1.2 to 2. The stellar line should mainly be caused by B stars, and primarily depends on the age of the starburst. Several of our targets will certainly have a contribution from B stars , and for those cases the $f_{esc,CII}$ will be underestimated. On the other hand, winds can work in the opposite direction. An out-flowing ISM can create a broad blue wing, where the line center appears less optical thick than it would be without winds. However, measuring the interstellar lines for a large sample FUSE galaxies (including some in this paper), \citet{2009ApJS..181..272G} found that in all cases the out-flowing neutral gas has much lower density than at the line core. We therefore estimate that this effect will be no larger than the effect from the stellar photospheric absorption.   In addition, the method does not allow for any dust within the holes, and from all these effects combined it is evident that $f_{esc,CII}$ only serves as an indicator for the escape fraction we would derive if we could measure directly on the LyC. With this in mind, we will use  $f_{esc,CII}$ together with $f_{esc}$ in the next section, to evaluate how the escape fraction correlates with other properties of the galaxies.

In Fig.~\ref{figure:multispec} the intrinsic \ion{C}{ii} $\lambda$1036 {\AA} lines in the FUSE spectra are plotted for the whole sample (except VV 114). Also, the intrinsic Ly$\beta$ and, where possible, the Milky Way  \ion{C}{ii} $\lambda$1036 {\AA} lines are shown. The latter can be used as a zero flux level calibrator if plotted with high resolution, and was found to agree well for all our targets. The f$_{esc,rel,CII}$ ratio was measured in both the LiF 1A (which has the best S/N) and LiF 2B spectra, and the agreement between the two is mirrored in the errors. In addition, the agreement  for the same spectra fully reduced by CalFUSE is included in the errors. The pipeline background problem is not as severe at these longer wavelengths, and these spectra give a valid sanity check to our data.  If an agreement was not found, only upper limits are stated. The final errors are the combined errors of the statistics and of the range of results from measuring on the different spectra.  We do not include uncertainties caused by systematic errors like underlying absorption, winds or model limitations. 

For SBS 0335-052 we make an exception about the LiF1A and LiF2B agreement. The best S/N spectrum LiF1A clearly shows an unsaturated line (see Fig.~\ref{figure:multispec}), and we measure a value close to that derived in \citet{2009ApJS..181..272G} and \citet{2011ApJ...730....5H}.  We note however that the more noisy LiF2B spectrum, in one single pixel shows zero flux in the line center. This disagreement is not included in the error, which is still large due to the very narrow line.  We also note the low metallicity of this galaxy (12+log(O/H)=7.3), which possibly could affect the estimated escape fraction if the abundance of carbon is very low.

 For the higher redshift targets, the measurements were performed on the LiF 2A (MRK 0357) and SiC 2B (Mrk 54) detector segments. For Tol 1247-232 the  \ion{C}{ii} $\lambda$1036 {\AA} line falls in the gap between the two LiF detector segments, and the quality of the SiC 2B spectrum plotted in Fig.~\ref{figure:multispec} is not sufficient to allow for any measurements. 

For the whole sample in Table~\ref{table:lyman} (excluding SBS 0335-052), the upper limit to the mean escape fraction is $<$$f_{esc,CII}$$>$ $<$ 2.7 $\%$ . The  value for the same sample of galaxies that were included in the stack (filled rows in column 2, Table~\ref{table:lyman}) is  $<$$f_{esc,CII}$$>$ $<$ 2.1$\%$. 

One galaxy stands out in the sample, ESO 338-IG04, with $f_{esc,CII}$ = 16.2$\pm$1.3 $\%$. Unfortunately this is one of the more nearby galaxies, and with the Lyman limit redshifted to only 920 {\AA}, we cannot measure the LyC flux between the Milky Way+airglow Lyman series below this.

\section{Properties connected to $f_{esc}$ }
\label{corr}
With this small sample of local star forming galaxies, we take a tentative look for hints about what physical properties might regulate the escape of ionizing photons. Some of these properties were also investigated in the cosmological SPH simulations of several hundreds star forming galaxies in \citet{2011MNRAS.412..411Y}, which we will use for comparison. We do not expect tight correlations since the conditions for leakage depends on so many parameters, what we hope to find are indications of trends.  Here, we investigate how $f_{esc}$ is related to dust absorption, stellar and \ion{H}{i} mass, metallicity, Ly$\alpha$,  and specific star formation rate. Other factors like the morphology and the aspect  angle are also likely to be important for the escape of ionizing photons, but are beyond the scope of this paper. 

  It is important to keep in mind for the interpretation of  Fig.~\ref{figure:corr}, that a significant residual intensity in the core of the \ion{C}{ii} $\lambda$1036 {\AA} absorption line is a necessary but not sufficient condition for the escape of ionizing radiation \citep{2011ApJ...730....5H}. Thus, the two galaxies that are contributing the strongest to the trends (ESO338-IG04 and SBS 0335-052) might not be giving the whole truth. This is especially true for SBS 0335-052 as discussed in the previous section, and if the escape fraction estimate for this galaxy is faulty several of the trends in Fig~\ref{figure:corr} will be weaker or even non-existent.
 
 In Fig~\ref{figure:corr}, the rhombs symbolizes the galaxies for which $f_{esc}$ was measured on the Lyman continuum, and the triangles are the galaxies where $f_{esc,CII}$ was estimated from the residual \ion{C}{ii} $\lambda$1036 {\AA} flux. Filled symbols represents galaxies with Ly$\alpha$ in emission, and open symbols galaxies with Ly$\alpha$ in absorption.\\

\noindent \large{\textit{$f_{esc}$ vs dust absorption}}\\ \normalsize
In Fig~\ref{figure:corr}a, the escape fraction is plotted against the internal absorption, E(B-V)$_i$. For Tol 1247-232 we have added a hatched line indicating the values $f_{esc}$ would take depending on the correct treatment of dust absorption, starting from what we measure from the IUE spectrum (E(B-V)$_i$=0.21) and down to zero absorption as indicated by the FUV slope in the FUSE spectrum \citep{2002A&A...393...33B}. 

The most interesting result of this plot is perhaps that no galaxies with high intrinsic absorption show any sign of LyC leakage. There might also be a trend for a higher $f_{esc}$ with a lower internal absorption, which also has been observed for the escape fraction of Ly$\alpha$ photons (e.g. \citealt{2009A&A...506L...1A, 2010Natur.464..562H, 2011ApJ...730....8H}), although the escape mechanisms  can vary significantly between the two. In \citet{2011MNRAS.412..411Y} there was a significant effect of dust attenuation on $f_{esc}$, but only a small effect was seen in the models by  \citet{2008ApJ...672..765G} and \citet{2010ApJ...710.1239R}. \\
\\

\noindent \large{\textit{$f_{esc}$ vs metallicity}} \\ \normalsize
In Fig~\ref{figure:corr}b we have plotted $f_{esc}$ against the gas phase metallicity given by the oxygen abundance. There is no obvious trend in the data, but possibly it could be said that a larger fraction of the low metallicity galaxies shows sign of leakage than of the high metallicity galaxies. 

The metallicities have been obtained from the literature, and are usually derived for the central regions of the galaxies. However, unlike dust, it has been shown for local interacting galaxies that the metallicity is quite stable over the whole galaxy (e.g. \citealt{2002A&A...390..891B,2010ApJ...723.1255R}). The metallicity is not expected to behave as the dust, and the two are not correlated in the local starbursts studied in \citet{1994ApJ...429..582C}, where the data are consistent with a unique extinction law independent of metallicity.  

\citet{2011MNRAS.412..411Y} found a negative correlation between the escape fraction and the metallicity, which they argue  is due to the positive correlation between halo mass and metallicity. The similarity between Fig~\ref{figure:corr}b and c,  is also more likely an effect of massive galaxies being more efficient in keeping the elements produced by the stars (the mass-metallicity relation, e.g. \citealt{2004ApJ...613..898T}), than an actual $f_{esc}$-metallicity relation. \\ 
\\






\begin{figure}[t!]
\centering
\includegraphics[width=9cm]{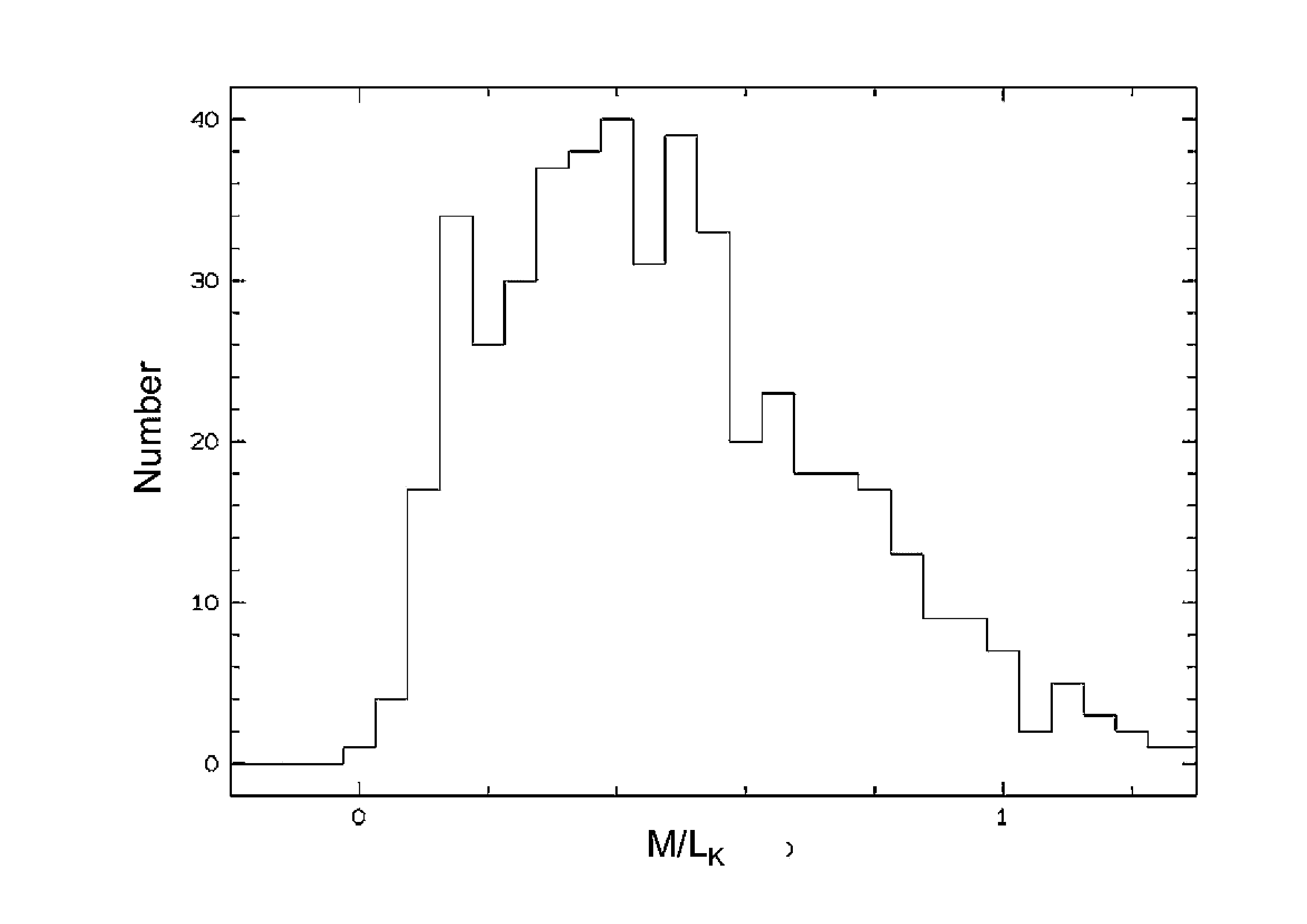}
\caption{The distribution of mass to light ratios (\mass/L$_K$) in solar units, for a sample of star forming SDSS galaxies. The selection criteria were EW(H$\alpha$)$>$100 {\AA}, -18 $<$ M$_R$ $<$ -23,  and z $>$ 0.1. The mean value is $<$\mass/L$_K$$>$=0.54, and median=0.48. The adopted value was  \mass/L$_K$=0.5.}
\label{figure:mlr}
\end{figure}

 \begin{figure*}[t!]
\centering
\includegraphics[width=9cm]{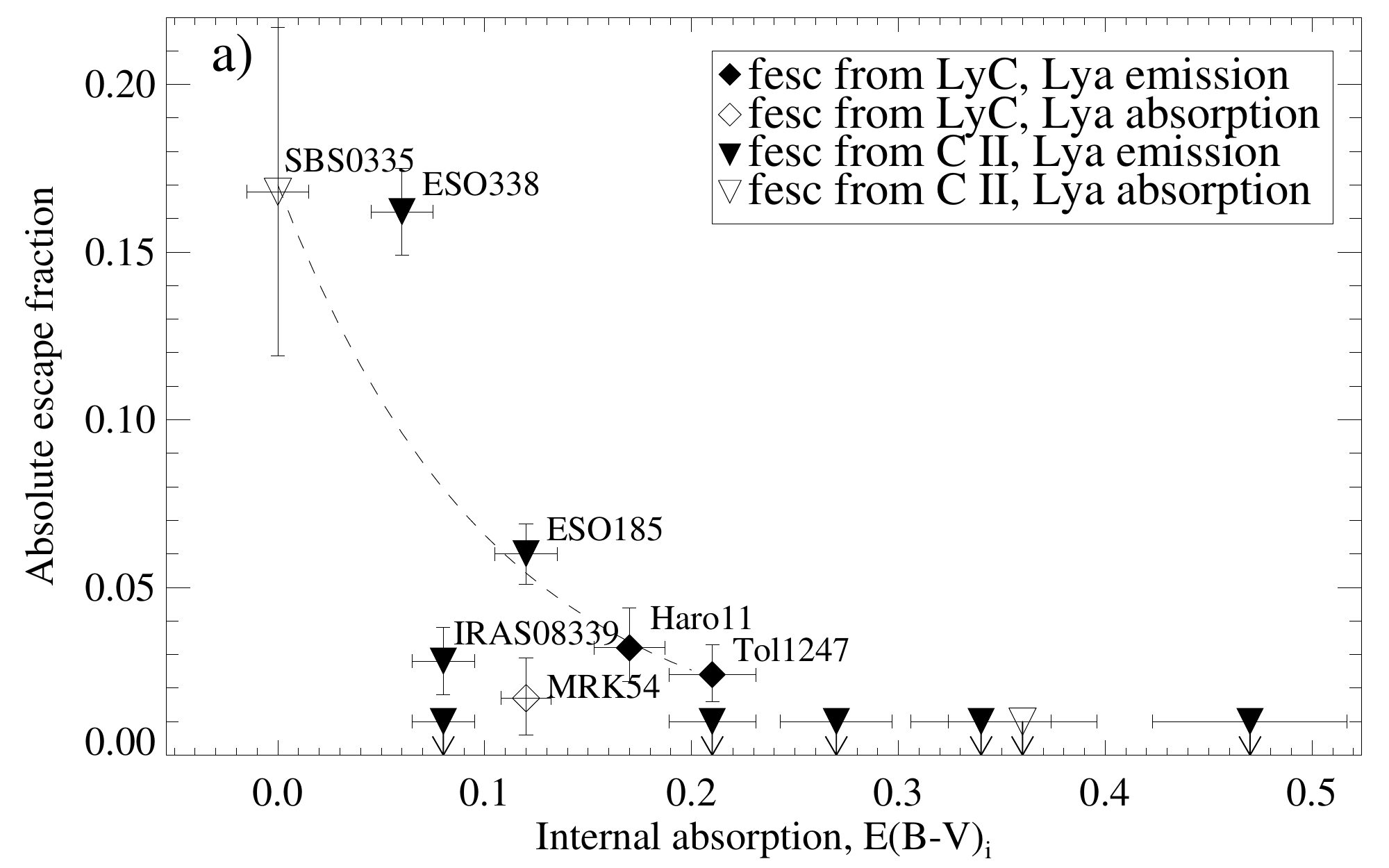}
\includegraphics[width=9cm]{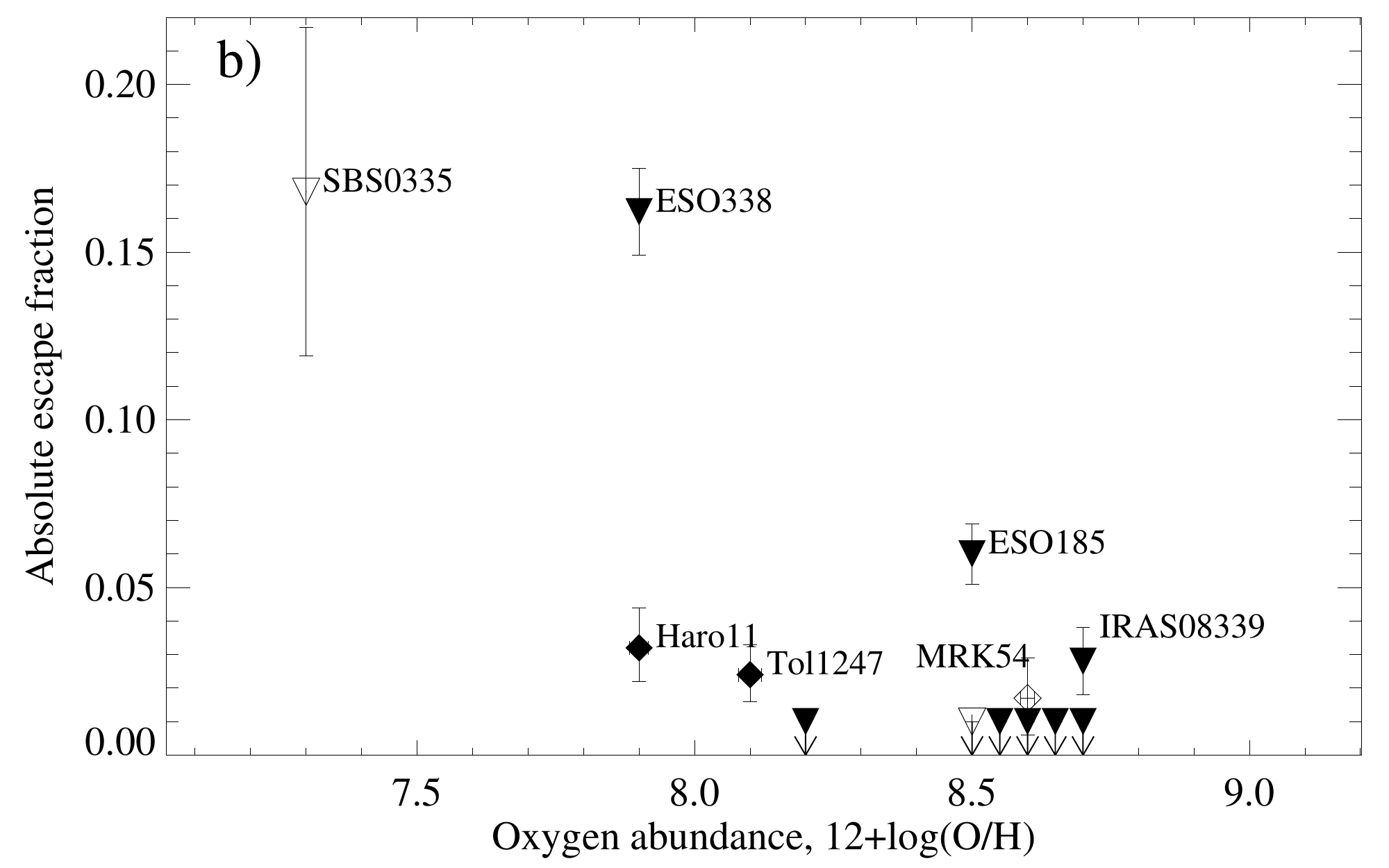}
\includegraphics[width=9cm]{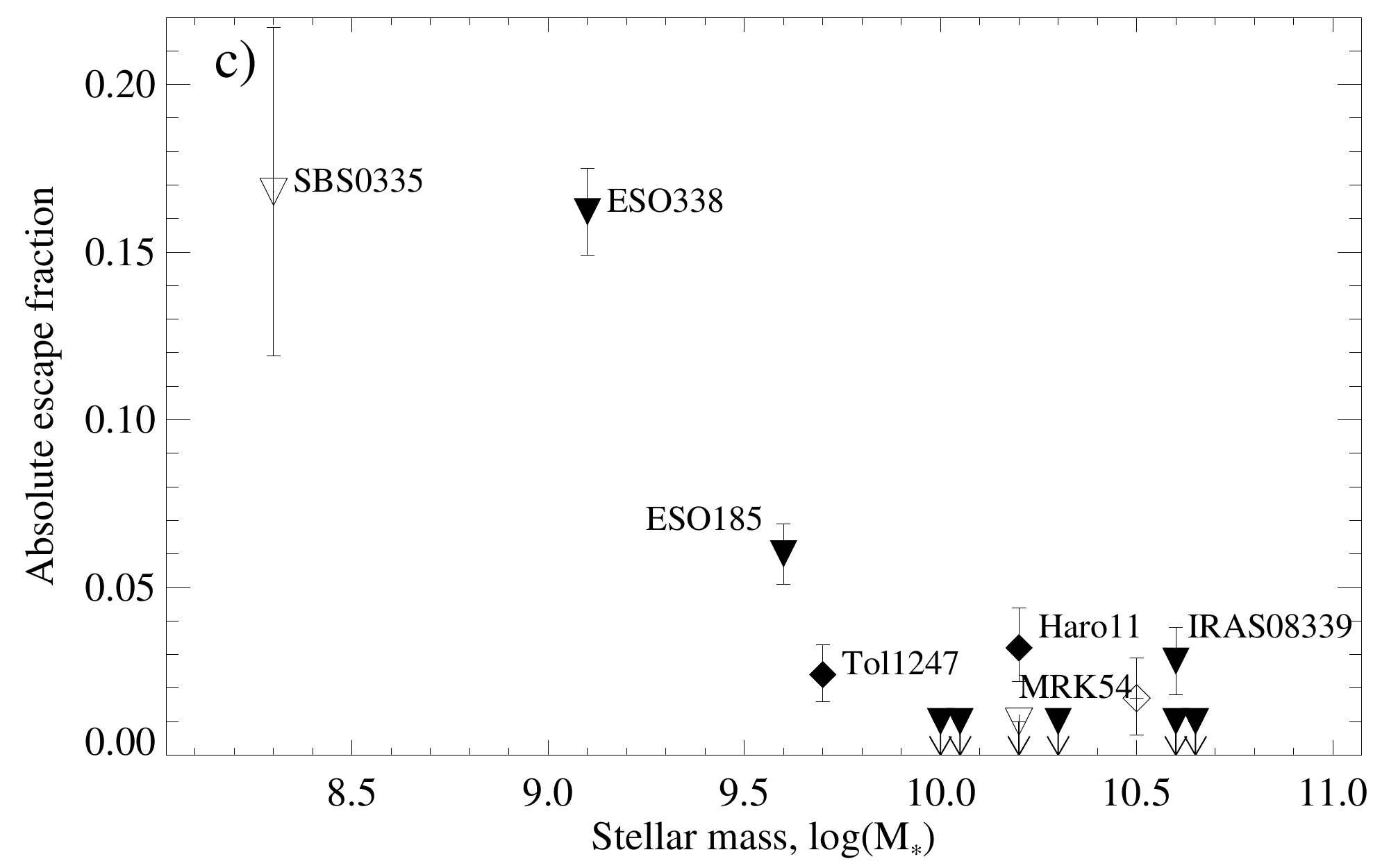}
\includegraphics[width=9cm]{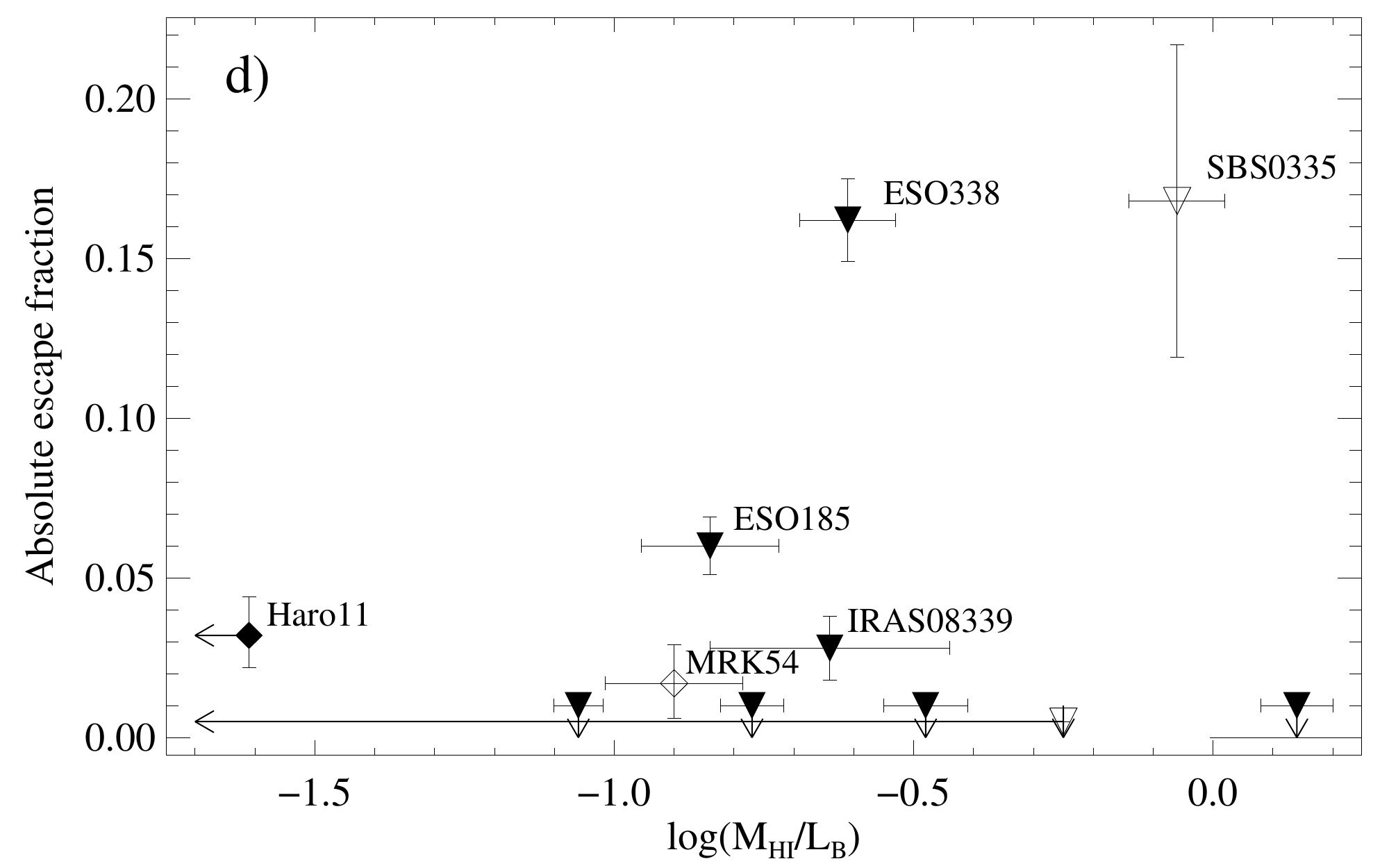}
\includegraphics[width=9cm]{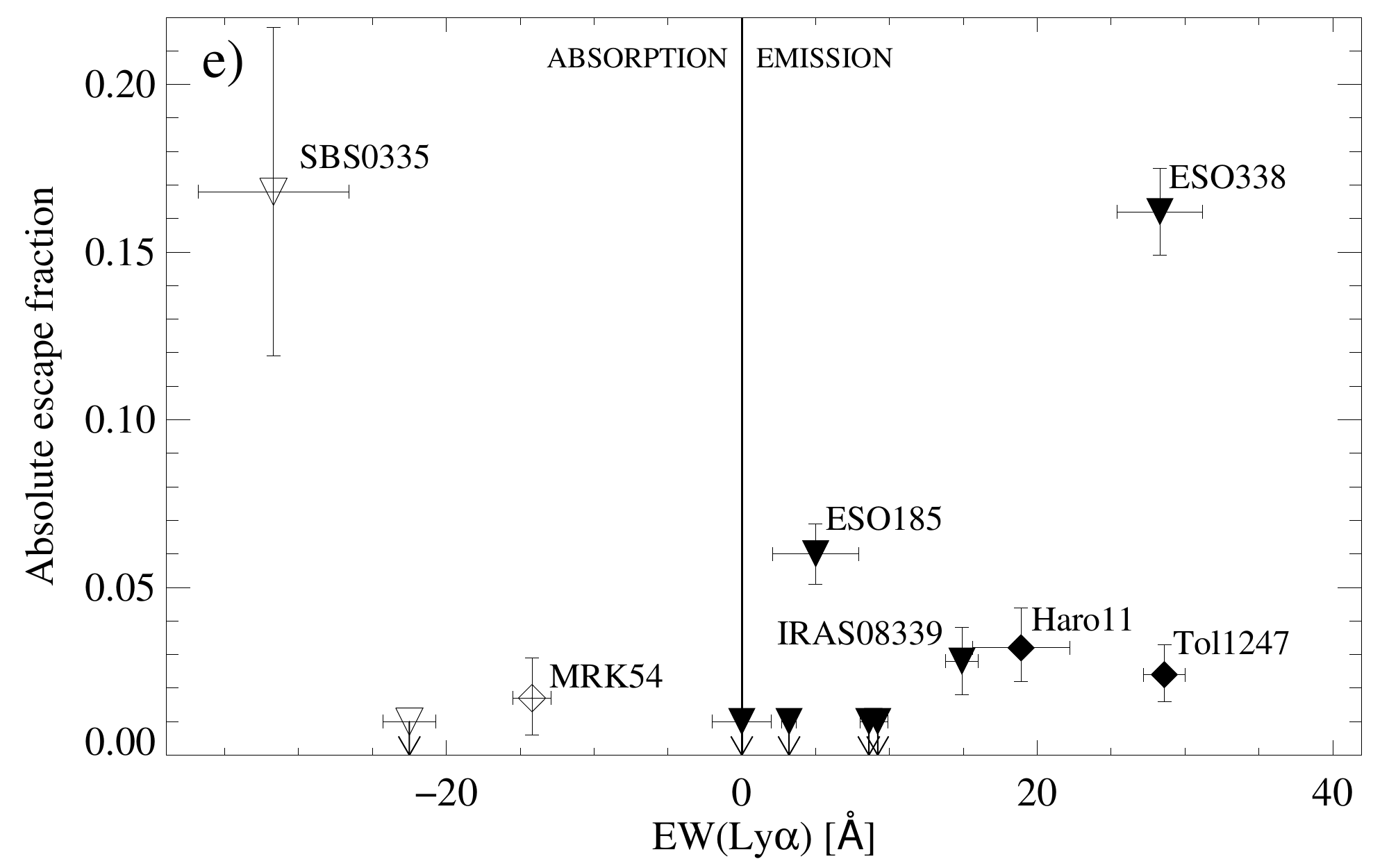}
\includegraphics[width=9cm]{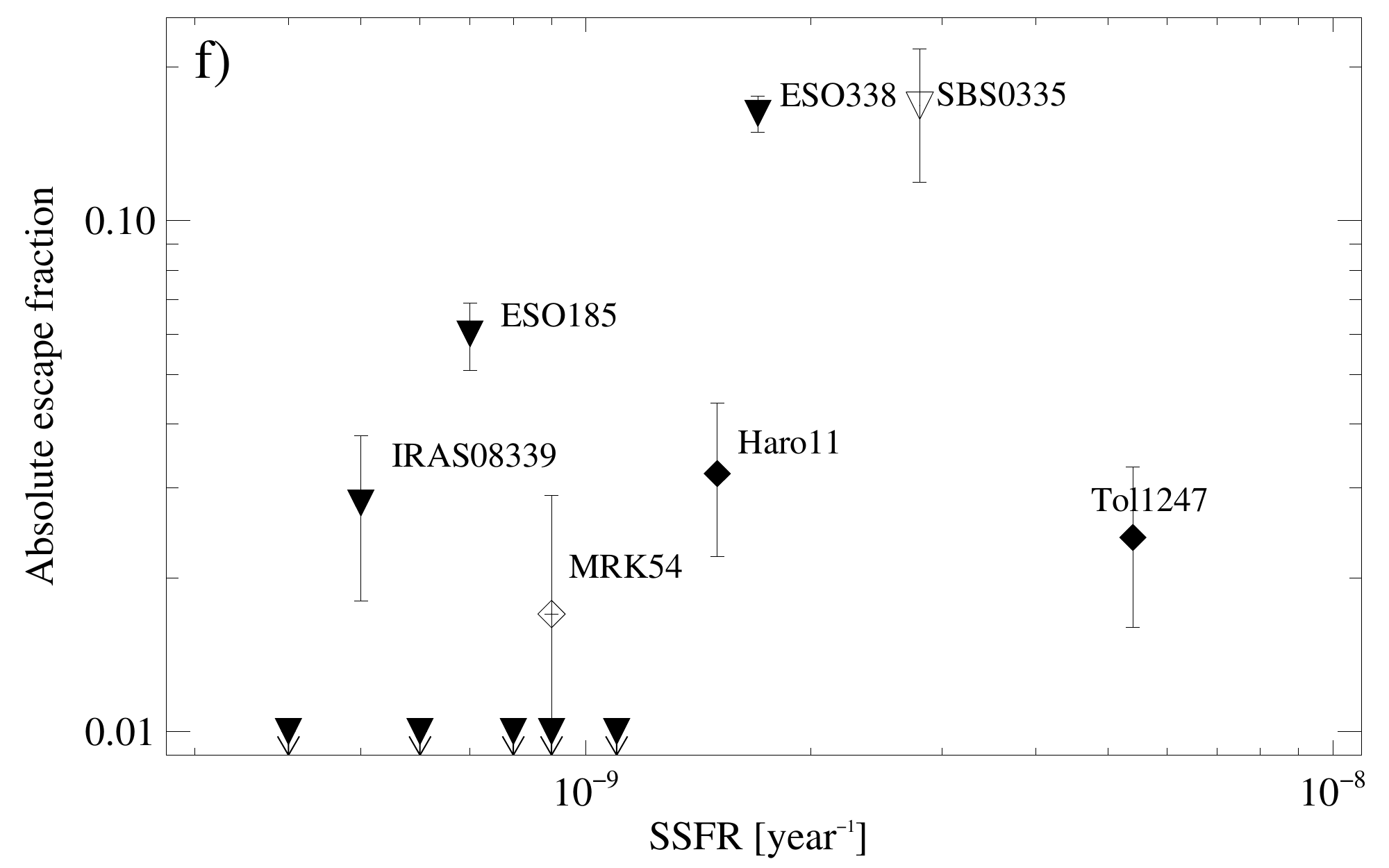}
\caption{A  tentative plot of the escape fraction plotted against various properties of the galaxies. The rhombs symbolize galaxies where $f_{esc}$ was measured directly on the Lyman continuum, the triangles galaxies where $f_{esc}$ was estimated using the residual flux in the \ion{C}{ii} $\lambda$1036{\AA} line. Filled symbols are plotted for galaxies with Ly$\alpha$ in emission, open symbols for galaxies with Ly$\alpha$ in absorption. The $f_{esc}$ errors are plotted for 1$\sigma$. An abridgment of the names of the most interesting galaxies are also written out.  In panel \textbf{a)} $f_{esc}$ against the intrinsic absorption, E(B-V)$_i$.  The hatched line shows the position of Tol 1247-232 for the range of E(B-V)$_i$=0--0.21 as described in the text, \textbf{b)} $f_{esc}$ against the oxygen abundance, 12+log(O/H), \textbf{c)} $f_{esc}$ against stellar mass [\sma], \textbf{d)}  $f_{esc}$ against \ion{H}{I} mass related to the number of ionizing photons via L$_B$ (solar units), 
\textbf{e)} $f_{esc}$ against Ly$\alpha$ equivalent width, and \textbf{f)} $f_{esc}$ against the  specific SFR (SFR/\mst). }
\label{figure:corr}
\end{figure*}

\noindent \large{\textit{$f_{esc}$ vs mass}} \\ \normalsize
Models  seem to indicate that  the galaxy mass is an important parameter in determining  the amount of ionizing photons that escapes. In most models  less massive haloes have higher escape fractions (e.g. \citealt{2010ApJ...710.1239R,2011MNRAS.412..411Y}), but others show  a higher escape fraction for more  massive disk galaxies \citep{2008ApJ...672..765G}.

To estimate the stellar mass (\mst) of the galaxies, a conversion of the mass to light ratio in the K band (\mass/L$_K$) was derived. For this purpose, we applied a model of the stellar population in galaxies from Bergvall et al. 2013 (in prep.). The model assumes that the stellar population is composed of a young,  star forming component, mixed with an old stellar population, with SEDs obtained from the \citet{2001A&A...375..814Z} models. The age and relative mass of the young component is constrained by the H$\alpha$ equivalent width and the spectral shape. Dust reddening is taken into account as derived from the H$\alpha$/H$\beta$ ratio, corrected for underlying absorption. The metallicities were assumed to be 20$\%$ solar.

The model was run on a sample of  star forming SDSS galaxies with redshifts z $>$ 0.1 (to include most of the light within the 3$\arcsec$ diameter spectroscopic aperture), luminosities matching our sample -18 $<$ M$_R$ $<$ -23, and  H$\alpha$ equivalent width EW(H$\alpha$) $>$ 100 {\AA}. The distribution of the SDSS galaxies can be seen in Fig.~\ref{figure:mlr}, with the mean value $<$\mass/L$_K$$>$=0.54, and median=0.48. We will adopt  \mass/L$_K$=0.5 in our estimate of the stellar masses (compare with \mass/L$_K$=0.4 found for spiral galaxies in \citealt{2001ApJ...550..212B}). 

The K band magnitudes were for most  galaxies obtained from the two Micron All Sky Survey (2MASS) data via the NASA/IPAC Extragalactic Database (NED). For the two galaxies lacking 2MASS data, the K magnitudes were obtained from \citet{2000A&A...363..493V} for SBS 0335-052, and \citet{1985RMxAA..11...91M} for Tol 1247-232. The derived stellar masses of the galaxies are listed in Table~\ref{table:lyman}. 
For three of the galaxies, Haro 11, ESO 338-IG04 and ESO 185-IG013, the stellar masses  were previously derived  using an independent method by \citet{2001A&A...374..800O}. Here, the photometric masses were derived by integrating the luminosity profiles for the disk and burst components, separately. The masses derived with our method agree very well for Haro 11, and the mean for all three galaxies within a factor 2.

In Fig~\ref{figure:corr}c,  $f_{esc}$ is plotted against the stellar mass. The plot seems to indicate what most models predict, that it is easier for ionizing photons to escape from galaxies with lower mass.  \\ 
\\

\noindent \large{\textit{$f_{esc}$ vs \ion{H}{I} mass} \\ \normalsize
In this plot the aim is to relate the hydrogen mass to the number of ionizing photons. To a first approximation this is given by the absolute B magnitude. The \ion{H}{i} masses were taken from a compilation by Bergvall et al. 2013 (in prep.), including also new data  (listed in Table~\ref{table:lyman}).  
 
  From Fig~\ref{figure:corr}d, it does not seem as if the hydrogen mass over B luminosity has any evident impact on the escape fraction. This could perhaps be interpreted as another evidence for escape via a clumpy interstellar medium, through channels, or by young star clusters that has been displaced from the main \ion{H}{I} halo. If the hydrogen gas was evenly distributed around the starburst, the \ion{H}{I} mass would certainly need to be very low in order for any leakage to take place. \\
\\

\noindent \large{\textit{$f_{esc}$ vs Ly$\alpha$}} \\ \normalsize
It is also interesting to investigate if LyC photon escape can be connected to the escape of Ly$\alpha$ photons.  It is not obvious that the two should be connected since the recombination line strengths are proportional to (1-$f_{esc}$) in the picket-fence model, but for low escape fractions the two could possible be related by common escape mechanisms. Due to the resonant nature of the Ly$\alpha$ transition,  Ly$\alpha$ photons can either escape through ionized or empty channels in the ISM, after multiple scatterings in a diffuse halo, or through an out-flowing medium where they will be Doppler shifted out of the line core. The LyC photon on the other hand, can only escape via the first of these alternatives.

\citet{2009AJ....138..923O} showed with high resolution imaging that the Ly$\alpha$ equivalent width is changing on very small scales, indicative of a clumpy ISM. If we have ionized or empty channels, both the Ly$\alpha$ and the LyC photons would have a clear path out, and Haro 11 and ESO 338-IG04 were both shown to have regions where the Ly$\alpha$/H$\alpha$ ratio follow the theoretical recombination case B. This could indicate a picket-fence scenario, but the same effect would also be seen if we have an out-flowing neutral ISM. If the Ly$\alpha$ photons are escaping through holes, we would expect a higher EW(Ly$\alpha$) together with a higher $f_{esc}$, while if the Ly$\alpha$ photon escape is via an out-flowing medium or in a scattered component, we would expect no relation.

 In Fig~\ref{figure:corr}e,  $f_{esc}$ is plotted against the Ly$\alpha$ equivalent width.   There is no obvious trend here, and LyC photons seems to be able to escape from both  Ly$\alpha$ emitters and absorbers, which might indeed indicate that LyC photons and the main part of the Ly$\alpha$ photons are escaping through different mechanisms.\\
 \\
 
  

\noindent \large{\textit{$f_{esc}$ vs specific SFR}} \\ \normalsize
 All local star forming galaxies studied so far have been found to be opaque to LyC photons if the gas is homogeneously distributed (e.g. \citealt{2001ApJ...558...56H, 2006A&A...448..513B, 2009ApJS..181..272G}). It seems like LyC photons therefore must escape through a porous interstellar medium, according to the so called picket-fence model. Such holes can be caused by feedback effects from star formation. A powerful starburst could in principle ionize cone-like paths out of the galaxy, or winds and supernovae (SN) explosions could create low density tunnels through which the ionizing photons can escape. A porous ISM is therefore closely related to the star formation, and the star formation rate per unit mass could potentially give an indication of how big impact the star formation will have on the host galaxy. 
 
The SFRs of the galaxies were estimated from the sum of SFR(UV) plus  SFR(IR), using the recipes in \citet{1998ARA&A..36..189K}. The UV flux density was measured in the Galactic extinction corrected IUE spectrum at 1500 {\AA}. The IR luminosity was integrated over the 
IR spectrum between 8 - 120 $\mu$m, with data obtained  from the Infrared Astronomical Satellite (IRAS) for all galaxies except one. For SBS 0335-052, the data were instead obtained from the Multiband Imaging Photometer (MIPS) onboard the Spitzer space telescope. 
 
In Fig~\ref{figure:corr}f, $f_{esc}$ is plotted against the SSFR (SFR normalized by stellar mass (SFR/\mst)). The result is difficult to interpret, mainly due to the outlier Tol 1247-232 which seems to be forming stars at an extraordinarily high rate. Possibly,  the plot can be interpreted as galaxies above a certain SSFR, here $\gtrsim$1.5 $\times$ 10$^{-9}$ yr$^{-1}$, all show signs of leakage, similar to the results found in \citet{2011MNRAS.412..411Y} (although these have been normalized by halo mass).  We note that most of the galaxies with high SSFR also have been documented as Wolf Rayet galaxies in the literature (Table~\ref{table:tfour}), implying young stars and energetic  mechanical feedback.

\subsection{Notes on individual targets}
\label{individ}
\noindent  \textit{Haro 11} \\
Haro 11 was the first local galaxy for which LyC leakage was detected (\citealt{2006A&A...448..513B}, L11). It is one of the most studied BCG galaxies in the local universe (e.g. \citealt{2002A&A...390..891B,2007ApJ...668..891G,2010MNRAS.407..870A,2010MNRAS.405.1203M}), and has been found to have low a low metallicity, a vigorous star formation, and a morphology reminiscent of a merger event  \citep{2001A&A...374..800O}. It was singled out by Bergvall to be observed by FUSE because of its surprisingly low \ion{H}{i} content (still undetected) for its type, and its extreme starburst properties, both favoring leakage. 

Haro 11 has long been known also to be a Ly$\alpha$ emitter  (e.g. \citealt{1998A&A...334...11K}). There is an intriguing region (knot C) that emits Ly$\alpha$ photons at Case B recombination values \citep{2007MNRAS.382.1465H}, which could indicate a leakage of LyC photons as well.  This escape scenario is strengthen by recent findings by Sandberg et al. 2013 (submitted), where integral field spectroscopy shows a weak inflow of neutral gas towards knot C. The escape of Ly$\alpha$  photons can thus not be connected to outflows as has otherwise been found to be a common escape mechanism,  in principle ruling out any other scenario than Ly$\alpha$ escape via a low covering fraction for this region.  \\
\\

\noindent  \textit{SBS 0335-052}\\
This low mass galaxy is interesting in many aspects. It is one of the most metal poor galaxies known in the local universe \citep{1999ApJ...511..639I}, and seems to have a negligible dust absorption.  It is a strong Ly$\alpha$ absorber, but has been found to emit Ly$\alpha$ on small scales outside the central region \citep{2009AJ....138..923O}. SBS 0335-052 seems to be an extremely young starburst (e.g. \citealt{2010ApJ...725.1620A}), possibly even so young that no winds from massive stars have had time to develop yet, which is interesting in the discussion about the importance of a porous ISM for escape.  

We want to caution about the uncertain data for this galaxy, where the two spectra with the  \ion{C}{ii} $\lambda$1036 {\AA} line was found to diverge profoundly. The escape fraction presented here was estimated to 17 $\%$ based on the best S/N spectrum, but the other spectrum would give zero flux in the line center and no escape. \\
\\

\noindent  \textit{IRAS 08339+6517}\\
This is a galaxy with spiral structure, containing a central starburst which  shows a central bright knot of Ly$\alpha$ emission \citep{2009AJ....138..923O}. IRAS 08339+6517 has a very low dust absorption, but is one of the most metal rich galaxies in the sample \citep{1998ApJ...495..698G}. It also stands out for showing signs of  LyC leakage, even though it is one of the most massive galaxies in the sample. Possibly, the fact that we are observing it in the polar direction from the disk (face-on) can explain the indication of LyC leakage, as found for modeled disk galaxies by \citet{2008ApJ...672..765G}.   \\
\\ 

\noindent  \textit{Tol 1247-232}\\
This is the second local galaxy for which LyC leakage has been detected, with $f_{esc}$ = 2.4$^{+0.9}_{-0.8}$ $\%$. Again, this might be a lower limit since there are indications of a much lower dust absorption from FUSE data \citep{2002A&A...393...33B}, and the escape fraction could be as high as 17 $\%$.  The question about the intrinsic absorption will be further investigated with new data from the HST, which will also allow for a derivation of the intrinsic ratio ($f_{1500\AA}/f_{900\AA})_\mathrm{int}$ following  the method with bin-wise SED modeling by \citet{2007MNRAS.382.1465H}.

Tol 1247-232 is an interesting galaxy in many other aspects as well. \citep{2002A&A...393...33B} found it difficult to fit this extremely blue galaxy with any SED model. Measuring the flux in the same spectral regions in  the FUSE spectrum and applying the \citet{2001A&A...375..814Z} model, we arrive at the same conclusion, that Tol 1247-232 is bluer than the youngest SED model.   Interesting perhaps to note, is that bluer-than-SED  colours also have been observed for several high redshift LyC leaking galaxies (e.g. \citealt{2009ApJ...692.1287I}). \citet{2011MNRAS.411.2336I} could fit these colours by including pop III stars in the model, but local galaxies containing such stars are yet to be found.

 \citet{2002A&A...393...33B} also found that the best fit implies a very recent burst (1 Myr) or a constant star formation over, at most, 5 Myr.  The young age seems to be confirmed by radio continuum observations by \citet{2007ApJ...654..226R},  with four out of five indices agreeing with a very young burst age. In the optical, the [\ion{O}{iii}] lines are strong, but the [\ion{N}{ii}] lines are weak relative H$\alpha$ \citep{1991A&AS...91..285T}, so it does not qualify as an AGN from BPT diagnostics (see Fig.~\ref{figure:bpt1}), but  a minor contribution from an AGN cannot be excluded. Also Wolf Rayet features in the optical imply ages $<$ 10 Myr \citep{1999A&AS..136...35S}. There seems to be a violent  star formation going on, and Tol 1247-232 is clearly off-set from the other galaxies  with SSFR = 5.4 $\times$10$^{-9}$ yr$^{-1}$.\\
 \\


\noindent  \textit{ESO 338-IG04}\\
This is again one of the galaxies observed with regions of direct Ly$\alpha$ emission in the \citet{2009AJ....138..923O} study. For one of the star forming knots, the Case B Ly$\alpha$/H$\alpha$ recombination value is even exceeded.   Also for this galaxy, Sandberg et al. 2013 (submitted), found indications for Ly$\alpha$ escape via a low covering fraction. 

ESO 338-IG04 fulfills the full parameter space for being a good LyC leakage aspirant, with little dust absorption, low stellar and \ion{H}{i} mass, and high SSFR. The  residual flux in the \ion{C}{ii} $\lambda$1036 {\AA} is also the most convincing case in the whole sample, and the derived escape fraction the highest with $f_{esc}$ $\sim$ 16 $\%$. 

In \citet{1985A&A...146..269B}, the optical/infrared colours of this galaxy were compared to predictions from a spectral evolutionary model including nebular emission. The best fit to the data indicated LyC leakage. Bergvall found that an unusually large fraction of the gas seemed to be ionized, which could explain the potential leakage. Later, the H I mass was determined to be 1.4{$\times$10$^9$\sma} (Bergvall et al. 2012, in prep.), which indeed is only 2-3 times larger than the H II mass. The H II mass is quite uncertain however and if the ionized medium is strongly filamentary or clumpy, it may be severely overestimated.\\ 
\\
\noindent  \textit{ESO185-IG013}\\
ESO 185-IG013 is displaying signs of being a LyC emitter,  both by estimating $f_{esc}$ directly from the LyC and from the \ion{C}{ii} $\lambda$1036 {\AA} line. The high resolution imaging by \citet{2011MNRAS.414.1793A} reveals hundreds of young stellar clusters with ages peaking at  3.5 Myr, and the morphology indicates that this  galaxy was involved in a recent merger event. ESO 185-IG013 is one of the least massive galaxies in the sample, but seems otherwise to be an average galaxy as compared  the rest of the sample.

\section{Discussion}
In order for galaxies to be able to re-ionize the universe, the escape fraction of ionizing photons would either need to be higher in the early universe (e.g. \citealt{2010ApJ...723..241S,2010ApJ...710.1239R}), or the faint-end slope of the luminosity function  steeper at high redshifts. The latter has recently been claimed by some  \citep{2009ApJ...692..778R,2010ApJ...709L..16O,2011arXiv1105.2038B} but is questioned by others \citep{2010MNRAS.403..960M,2011A&A...532A..33G}. An evolution of  $f_{esc}$ from high to intermediate redshifts  cannot be ruled out  as given by most observations over the last decade, but little is known about the escape fraction in the local universe. The stacked sample in this article is the first attempt for a unified estimate of the $f_{esc}$ parameter for local galaxies, although important selection biases is likely to be present in the sample constituting of 9 individually selected targets.  The result, $f_{esc}$ = 1.4$^{+0.6}_{-0.5}$ $\%$, is in accordance  with an evolving escape fraction as compared to most z $\sim$ 3-3.5 surveys (\citealt{2011ApJ...736...18N,2009ApJ...692.1287I,2006ApJ...651..688S,2001ApJ...546..665S}).  There are however other reports where upper limits of only a few percents have been claimed at these redshifts (e.g. \citealt{2011ApJ...736...41B,2010ApJ...725.1011V}). In contrast, these results combined with the stacked samples at z $\sim$ 1 where upper limits of 1-2 $\%$ have been found \citep{2010ApJ...720..465B,2010ApJ...723..241S,2007ApJ...668...62S,2009ApJ...692.1476C}, and the stacked spectrum of local galaxies in this work, seems to imply very little, or even no, evolution in $f_{esc}$  between z = 3 and 0. The question of an evolving escape fraction should perhaps thus still be regarded as an open question.

There is an aspect to consider that distinguishes the z $\sim$ 1 investigations to that of the local galaxies. Despite the larger number of galaxies observed at intermediate redshifts, no single detection has been made. In the local universe however, there have now been two detections of leaking galaxies. One of these was reported in this paper,  Tol 1247-232, for which we derive  an escape fraction of 2.4$^{+0.9}_{-0.8}$ $\%$. Also, at least one more galaxy, ESO 338-IG04, show signs of leakage as indicated by the residual flux in the \ion{C}{ii} $\lambda$1036 {\AA} line (see also \citealt{2009ApJS..181..272G,2011ApJ...730....5H}). It seems unlikely that there would be a gap around z $\sim$ 1 where no galaxies are leaking at a few percent or more, so the explanation could perhaps lie elsewhere. One possible explanation is the difference in at what wavelengths  the LyC is sampled. In all articles where z $\sim$ 1 galaxies have been observed, the Lyman continuum is sampled well below the Lyman limit (even down to 200 {\AA}  below) due to technical constraints, while both the local and the z$\sim$3 galaxies usually are sampled just below 912 {\AA}. This may perhaps be an indication of  that  the internal dust absorption in galaxies behaves in an unexpected way at these wavelengths, and usually the derivations of the extinction curve for galaxies ends at the Lyman limit (e.g. \citealt{2005ApJ...630..355C}).  It could also indicate that an idealized picket-fence model is not applicable, and perhaps even a small amount of dust in the holes can explain the non-detections when the Lyman continuum is sampled further from the Lyman limit, as the cross-section of dust to UV photons is predicted to rise rapidly at shorter wavelengths. It is also interesting to note that in this sample, the two galaxies with detections and modest escape fractions (Haro 11 and Tol 1247-232) both have a normal dust attenuation, while the two galaxies for which we have an indication of high leakage (ESO 338-IG04 and SBS 0335-052) both have a very low or negligible dust reddening as estimated by the IUE UV slope. If this trend is reliable, it seems likely that there exist a certain amount of dust also in the ionized channels, and it is only for galaxies with a very low dust content that the ionizing photons will survive out into the IGM. 


Some recent observations seems in favor of a steep faint-end slope of the luminosity function, and these galaxies have been suggested as the main source of ionizing photons in the early stages of the cosmic reionization. Most models also predict that low-mass galaxies should be much more efficient in letting out the ionizing photons into the intergalactic medium than more massive galaxies \citep{2000ApJ...542..548R,2010ApJ...710.1239R,2011MNRAS.412..411Y}. However, \citet{2008ApJ...672..765G} found a higher $f_{esc}$  for more massive disk galaxies at favorable sight-lines near the galactic poles.  The small  sample local galaxies studied here mostly lack a prominent disk, and for these there do seem to be a trend for a higher  $f_{esc}$ with lower mass, in favor of the former models.

In the local universe, no galaxies have been found to be transparent to ionizing photons if the gas is homogeneously distributed, rather it seems like a clumpy ISM is a necessary condition for leakage. In \citet{2002MNRAS.337.1299C} the critical star formation rate (SFR$_{crit}$) needed to form a patchy ISM  was derived. They find that the infall timescale of replenishing  neutral gas in z $\approx$ 3 LBG galaxies is short relative to the supernovae lifetimes, thereby making it possible to maintain a high SFR over a long time and form a highly porous ISM. In starburst galaxies in the local universe, the infall timescale (infall time per mass unit) is so high that the continuous replenishment of neutral gas into the star forming regions can prevent the porosity from ever attaining high values, even if the SFR {\it occasionally}, is high. In more massive galaxies, where the starburst activity is contained within the $\sim$ 1 kpc central dark matter potential well, the infall timescale may locally be extremely low. This would again prevent a high porosity to develop even if the SFR is extremely high. It might be, in the local universe, that only galaxies with spatially extended star forming activity and strongly enhanced specific SFRs can develop a porous ISM and allow a significant LyC escape, but if so, only over short periods. The relatively low escape fraction for the stacked sample of the local galaxies  found in this paper, compared to the usually higher values obtained for z $\approx$ 3 galaxies, might strengthen the arguments about the infall time scales and the formation of a porous ISM,  in theory working in favor of an evolving escape fraction.

\section{Conclusions}
Little is known about the escape fraction of hydrogen ionizing photons from local galaxies.  In this paper we utilize an improved  background subtraction routine developed in a previous paper, to investigate  LyC leakage from a sample of local galaxies in the FUSE  archive. In addition, the escape fraction was estimated from residual flux in the low ionization interstellar \ion{C}{ii} $\lambda$1036 {\AA} line. These are the main conclusions:

\begin{itemize}
\item  
One single detection of hydrogen ionizing leakage  was found measuring directly on the Lyman continuum. For the extremely young starburst Tol 1247-232, we derive an absolute escape fraction of $f_{esc}$ = 2.4$^{+0.9}_{-0.8}$ $\%$. The inconsistency between the dust absorption derived from the IUE data (E(B-V)$_i$=0.21) and FUSE data (E(B-V)$_i$$\approx$0),  could in principle allow an escape fraction of up to 17.4$^{+7.7}_{-6.7}$ $\%$.
\\
\item 
In the  derivation of the absolute escape fraction, the previously known LyC leaking galaxy Haro 11 was used as a calibrator to the intrinsic ratio, ($f_{1500\AA}/f_{900\AA})_\mathrm{int}$. For the first time, this ratio could be derived based on SED fitting from both continuum and emission lines.  The ratio was found to be ($f_{1500\AA}/f_{900\AA})_\mathrm{int}$ = 1.5$^{+0.6}_{-0.5}$ (or 4.2$^{+1.7}_{-1.4}$ in f$_{\nu}$).
\\
\item 
For the stacked sample of nine galaxies with mean redshift $<$z$>$=0.03, an excess in the   Lyman continuum with $f_{esc}$ = 1.4$^{+0.6}_{-0.5}$ $\%$ was measured. We caution however that  these individually selected targets might not be representative for typical star forming galaxies in the local universe. \\
\item 
From measuring the residual flux in the \ion{C}{ii} $\lambda$1036 {\AA} line and extrapolating out to the Lyman continuum, we found an upper limit of the mean escape fraction of the whole sample of $<$$f_{esc,CII}$$>$  $<$ 2.7 $\%$.
\\
\item 
Using the \ion{C}{ii} $\lambda$1036 {\AA} line as a tracer, an extraordinarily high escape fraction of  $f_{esc,CII}$ = 16.2$\pm$1.3 $\%$ was found for the galaxy  ESO 338-IG04. 
\\
\item
 A tentative investigation on  the behavior of  $f_{esc}$ versus other physical parameters was conducted. It is important to note that the escape fraction derived from the  \ion{C}{ii} $\lambda$1036 {\AA} line is highly uncertain, as it is a necessary but not sufficient condition for leakage to have a residual flux in the line core: 
\subitem
$\bullet$ No signs of LyC leakage was observed for galaxies with a high intrinsic dust absorption,  and there is a possible correlation for a  higher $f_{esc}$ at lower E(B-V)$_i$. 
\subitem
 $\bullet$ There is a possible trend for higher escape fractions for galaxies with lower masses.
\subitem
$\bullet$  A high specific SFR (SFR/\mst) seems to increase the chance of ionizing photons to escape. 


\end{itemize}

\begin{acknowledgements} 
E.L. gratefully acknowledges financial support from the Swedish National Space Board. N.B. and E.Z. also acknowledges financial support from the Swedish National Space Board and the Swedish Research Council. M.H. received support from Agence Nationale de la Recherche (reference ANR-09-BLAN-0234-01).\\
Some of the data presented in this paper were obtained from the Multimission Archive at the Space Telescope Science Institute (MAST). STScI is operated by the Association of Universities for Research in Astronomy, Inc., under NASA contract NAS5-26555. Support for MAST for non-HST data is provided by the NASA Office of Space Science via grant NAG5-7584 and by other grants and contracts. \\
This research has made use of the NASA/IPAC Extragalactic Database (NED) which is operated by the Jet Propulsion Laboratory, California Institute of Technology, under contract with the National Aeronautics and Space Administration. 
\end{acknowledgements}

\bibliographystyle{aa} 
\bibliography{sample2} 

\end{document}